\documentclass[a4paper,11pt]{article}
\usepackage{jheppub} 

\usepackage{xcolor}
\usepackage{graphicx}
\usepackage{braket}
\usepackage{subcaption}
\usepackage[normalem]{ulem}
\usepackage{amsfonts}
\usepackage{subcaption}

\usepackage{physics}
\usepackage{url}
\usepackage{tikz}
\usepackage{pgfplots}
\usetikzlibrary{fadings}
\pgfplotsset{compat=1.18}

\newcommand{\re}{\mathrm{Re}}
\newcommand{\Q}{\mathcal{Q}}

\newcommand{\N}{\mathcal{N}}

\title{Instanton condensation and a new phase of BPS black holes}

\author{Jack Holden}
\affiliation{Yau Mathematical Sciences Center, Tsinghua University, Beijing, 100084, China}
\emailAdd{jholden@tsinghua.edu.cn}

\abstract{
    We analyse the 1/16-BPS superconformal index for BPS black holes at equal charge in $AdS_5 \times S_5$, uncovering evidence for a new instability in the microcanonical ensemble along the small black hole saddle. This is indicated by instanton condensation in the matrix model description of the index. This instability occurs for black holes of radius close to, but below, the scale at which black holes become `small', and implies a new dominant phase in this region. We propose a connection to the partially deconfined phase in the field theory dual description. This would resolve recent confusion about the location of the partially deconfined phase in the BPS phase diagram and promises new avenues for understanding confinement, partial deconfinement, and the encoding of colour degrees of freedom under the holographic map. We also motivate the importance of instantons in partial deconfinement from a matrix model perspective.
}

\begin{document}

\maketitle
\flushbottom

\section{Introduction}

The description of black holes within the AdS/CFT correspondence has provided a powerful framework for uncovering new aspects of strongly coupled quantum field theories and quantum gravity \cite{Maldacena_1999, Witten:1998zw}. Understanding how black hole microstates are encoded in gauge theories offers a route by which the microscopic origin of black hole entropy and the organisation of degrees of freedom under the holographic map might be understood. By employing the duality in the contrary direction, meanwhile, gravitational solutions can be used to gain insight into strongly coupled gauge theories. 

A concept related to both aspects is the partially deconfined phase, a phase of gauge theories in which only a subset of colour degrees of freedom participates in confinement \cite{Hanada:2018zxn}. It was by exploring black hole thermodynamics that it was first appreciated that gauge theories, including QCD, generically exhibit this important phase \cite{Hanada:2016pwv}. Motivated by this connection, we employ the superconformal index to investigate the structure of the BPS black hole phase diagram, seeking signatures of the partially deconfined phase. We find evidence for a thermodynamic instability of the conventional BPS black holes at small charge, signalled by instanton condensation. We show that the new instability naturally invites interpretation as signalling the onset of the partial deconfinement.

Despite the success of partial deconfinement in explaining many aspects of black hole phase diagrams, its full representation in holography has some crucial unresolved questions. Our focus will be on the background $AdS_5 \times S^5$. In the thermal setting, it is well established that the $AdS$ vacuum and large black hole correspond respectively to the confined and deconfined phases. In contrast, the manifestation of the partially deconfined phase in the black hole phase diagram is not so transparent. Arguments have been proposed that all small black holes, defined as those with negative heat capacity, describe partially deconfined phases \cite{Hanada:2018zxn,Hanada:2016pwv}. An alternative suggestion is that only black holes that are sufficiently small as to be localised on the compact $S^5$ (via a Gregory-Laflamme instability \cite{Gregory:1993vy,Gregory:1994bj}) are partially deconfined, with this GL transition marking the onset of partial deconfinement.

To attack this problem, we will employ the superconformal index. The index allows some information about the strongly-coupled field theory to be computed using weak-coupling field theory techniques, in particular allowing the partition function and entropy of the theory at strong coupling to be approximated. Since black holes are identified with strongly-coupled field theories under holographic duality, the index has been employed successfully to compute quantities such as the black hole entropy from the quantum field theory. The index will therefore grant us the ability to perform a detailed analysis of the phase diagram of BPS black holes. We expect that the insights gained from the BPS black hole phase diagram will also inform our understanding of the thermal black hole phases.

Previous investigations offered no indication of the partially deconfined phase in the BPS black hole phase diagram. In fact, the supposed signal of the transition between complete and partial deconfinement, the Gross-Witten-Wadia (GWW) point, was located close to the black hole/string transition, contradicting physical expectations. We therefore set out to reinvestigate this black hole phase diagram. In particular, we explored matrix instanton contributions to the index, checking for instabilities against instanton condensation in the black hole phase diagram that had been established up until now. 

We provide strong evidence that instanton condensation does indeed occur, and that a new phase must dominate the microcanonical ensemble of the equal-charge BPS black hole phase diagram at sufficiently low charge. In particular, the new phase begins to dominate close to, but below, the charge at which the large and small black hole solutions meet, corresponding to an instability in black holes with radii just below the AdS scale. Moreover, we show that the purported new phase exhibits many of the properties we would expect of the partially deconfined phase. 

Exploring this aspect of the BPS black hole phase diagram is interesting for a number of reasons. First, and most directly, it would give evidence towards or against conjectured duals of the partially deconfined phase. This has major implications for understanding how colour degrees of freedom behave under duality, a central puzzle in holography. Second, it has the potential to enrich the black hole phase diagram. Several recent developments have argued for new, sometimes exotic phases in the black hole diagram or its field theory dual \cite{Choi:2024xnv,Gladden:2024ssb,Dias:2022eyq,Dias:2024edd,Choi:2021rxi,Cabo-Bizet:2019eaf}, with a comprehensive understanding of the full phase diagram still elusive. Third, we expect such study to inform us about the nature of confinement, deconfinement, and partial deconfinement, both in general and in the specific context of QCD. Indeed, as we will explain later, the critical role of instantons in partial deconfinement, along with implications for QCD and chiral symmetry breaking, was first realised by pursuing these questions about partial deconfinement and the superconformal index.

In addition, we offer a natural explanation for the unexpected placement of the GWW transition, which had previously caused confusion. We suggest the possibility of an unexpected Gregory-Laflamme instability for BPS black holes, and a possible route by which eigenvalues in the matrix model can be identified with nucleating D3 branes.

In Sec.~\ref{sec:pdec}, we review the partially deconfined phase. Sec.~\ref{sec:index} covers the superconformal index, its saddle point solutions, and how it is used to compute BPS black holes. Sec.~\ref{sec:truncations} explores saddle point solutions of the index as a matrix model in more detail
at particular truncations. Then, in Sec.~\ref{sec:numerical-results}, we present our results for the BPS black holes, arguing for the necessity of the multi-cut solutions. Finally, in Sec.~\ref{sec:discussion}, we discuss the interpretation of these results and possible future work.

\section{Partial deconfinement} \label{sec:pdec}

\begin{figure}  
    \centering
    \includegraphics[width=0.5\linewidth]{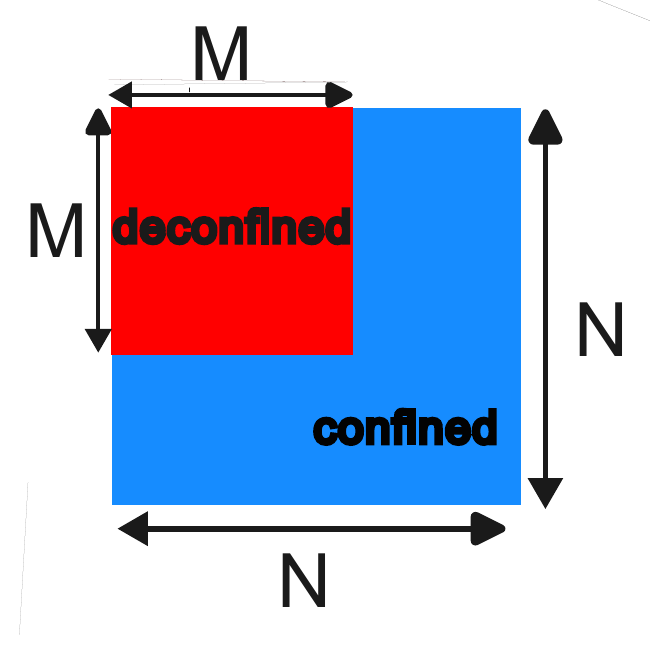}
    \caption{Cartoon depiction of the partially deconfined phase at large $N$. The colour degrees of freedom are split between $M$ confined and $N-M$ deconfined colours. There is effectively an $SU(N-M)$ confined sector and an $SU(M)$ deconfined sector.}
    \label{fig:pdec-cartoon}
\end{figure}

The partially deconfined phase is a phase of gauge theories intermediate between the confined and deconfined phases \cite{Hanada:2018zxn,Hanada:2016pwv, Hanada:2019czd, Hanada:2022wcq}. Specifically, the internal gauge group is split between confining and deconfining regions, as sketched in Fig.~\ref{fig:pdec-cartoon}. Below, we make this notion more precise, and explain the most crucial properties of the partially deconfined phase that will be relevant to the present work. 

\subsection{Polyakov loop eigenvalues}
To make the notion of `splitting of the gauge group' into confined and deconfined regions more precise, we will introduce Polyakov loop eigenvalues. A rigorous definition is easiest in the large $N$ limit, which we now assume. We will also assume zero chemical potential.

Recall that we introduce a temperature $T$ to a field theory by compactifying on a Euclidean $S^1$ with circumference $\beta=1/T$ (taking antiperiodic boundary conditions for fermions). We can probe the degrees of freedom of the gauge group by focussing on the eigenvalues of the holonomy $P$ around this $S^1$,
\begin{equation}
    P 
    \equiv
    \mathcal{P} \exp\left(\int_{S^1} \mathrm{d} t \,A_t\right)\, =     \begin{pmatrix}
    e^{i\alpha_0} &  &   \\
     & \ddots & \\
     &        & e^{i\alpha_{N-1}}
    \end{pmatrix}.
\label{eq:holonomy}
\end{equation}
where $\mathcal{P}$ denotes the path ordering. Taking the trace of the holonomy, $P$, gives us the value of the Polyakov loop, $\Tr~P$. We will refer to the phases $\alpha_i$ themselves as the eigenvalues for concision.
The set of all eigenvalues $\{ \alpha_i \}$ is gauge invariant up to permutation of the labels. In the large $N$ limit, we can represent this set by the density $\rho(\theta)$ of eigenvalues at each value of the phase $\theta$, 
\begin{align} \label{eq:rho-density-defn}
    \rho(\theta)=\frac{1}{ N} \sum_{i=1}^N \delta(\theta-\alpha_i)
    \implies
    \int_{-\pi}^{\pi} d \theta \rho(\theta) f(\theta) = \frac{1}{N}\sum_{i=1}^N f(\alpha_i)
\end{align}

\begin{figure} 
    \hfill\centering
    \begin{subfigure}[b]{0.3\textwidth}
    \centering
    \includegraphics[width=\linewidth]{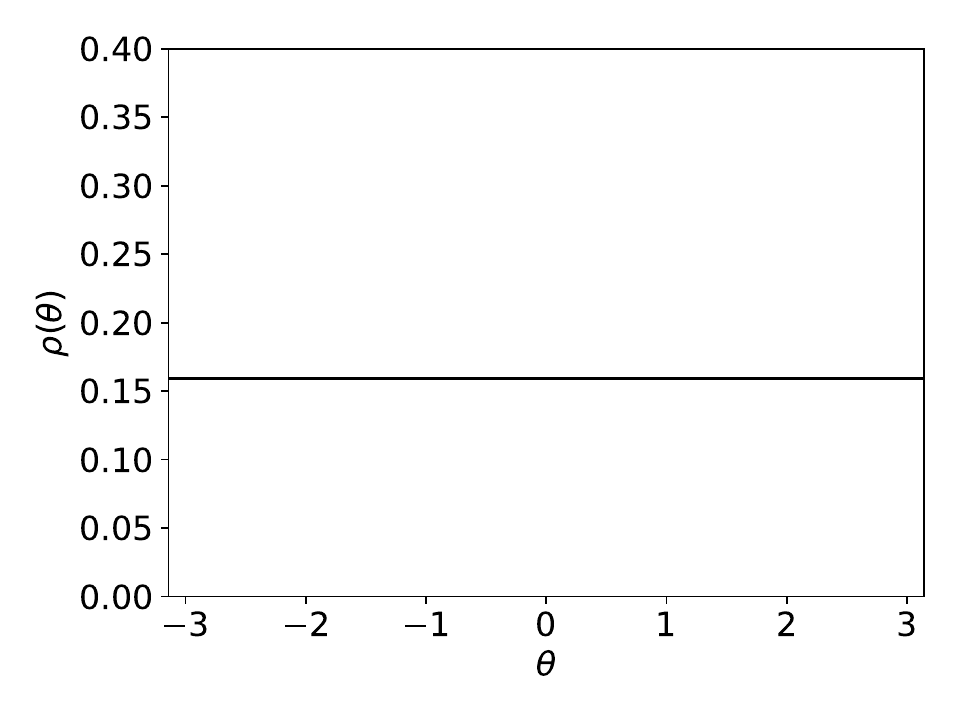}
    \caption{}
    \label{fig:gww-uniform}
    \end{subfigure}
    \hfill
    \centering
    \begin{subfigure}[b]{0.3\textwidth}
    \centering
    \includegraphics[width=\linewidth]{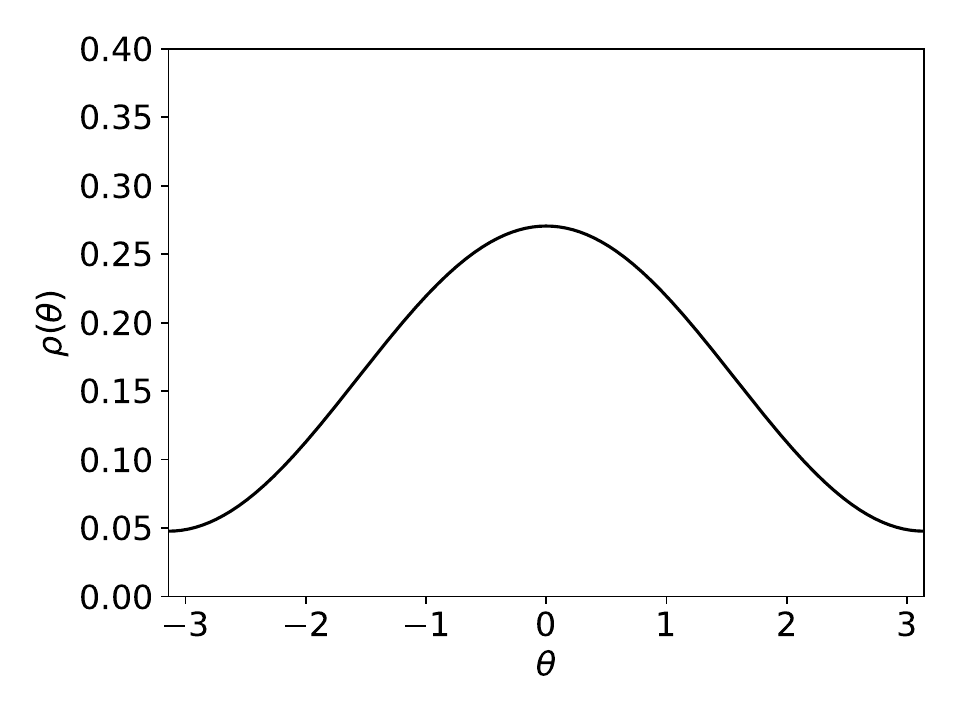}
    \caption{}
    \label{fig:gww-ungapped}
    \end{subfigure}
    \hfill
    \centering
    \begin{subfigure}[b]{0.3\textwidth}
    \centering
    \includegraphics[width=\linewidth]{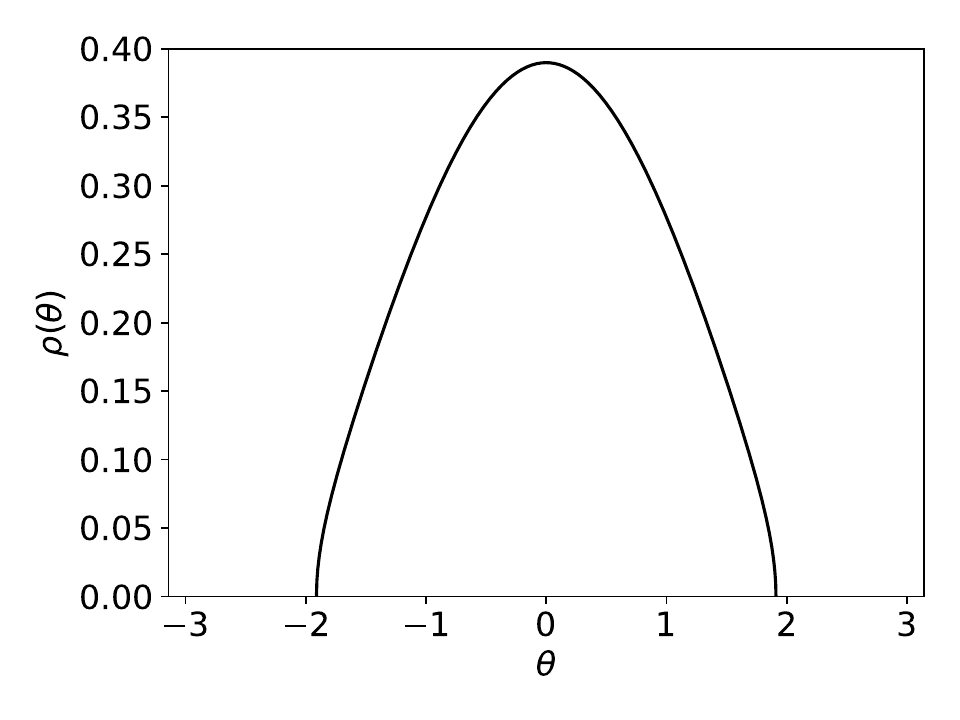}
    \caption{}
    \label{fig:gww-gapped}
    \end{subfigure}
    \caption{The three standard phases of large-$N$ gauge theories shown by Polyakov loop eigenvalue distributions, $\rho(\theta)$, at real coupling. At zero chemical potential, the uniform distribution (\subref{fig:gww-uniform}) corresponds to the confined phase. The distribution (\subref{fig:gww-gapped}) corresponds to the deconfined phase. The distribution (\subref{fig:gww-ungapped}) is described as \emph{non-uniform ungapped} and corresponds to the partially deconfined phase. At nonzero chemical potential, the interpretation of these distributions can change \cite{Hanada:2025rca}.} 
    \label{fig:gww-phases}
\end{figure}

It is well known that the Polyakov loop can be used as an order parameter for confinement and deconfinement. However, the distribution $\rho(\theta)$ offers a more informative description. In many physical situations, the distribution takes three distinct forms \cite{Gross:1980he,Wadia:2012fr}, shown in Fig.~\ref{fig:gww-phases}.  First is the uniform distribution, Fig.~\ref{fig:gww-uniform}, corresponding to the completely confined phase. The centre symmetry of the $SU(N)$, a signature of the confined phase, is seen by shifting the eigenvalue density distribution $\rho(\theta)=\rho((\theta+\delta) (\rm{mod~}2\pi))$. The deconfined phase is represented by a gapped phase, as shown in Fig.~\ref{fig:gww-gapped}. The centre symmetry is completely broken. Between these sits the non-uniform ungapped phase in Fig.~\ref{fig:gww-ungapped}. This is identified with the partially deconfined phase, where one can think in terms of an effective splitting $SU(N) \rightarrow SU(N-M)$ in the \emph{extended} (i.e. not gauge-invariant) Hilbert space. The confined part of the partially deconfined sector can be identified with the uniform component of the ungapped distribution, while the deconfined part corresponds to the rest of the density distribution. This division can be taken seriously, and, for example, it was argued in \cite{Gautam:2022exf} that the uniformly distributed component corresponds to a sector in colour space that produces linearly confining flux tubes, while the deconfined component does not. Note that the $Z_{N}$ centre symmetry is completely broken in this ungapped phase.

\begin{figure} 
    \centering
     \includegraphics[width=0.3\linewidth]{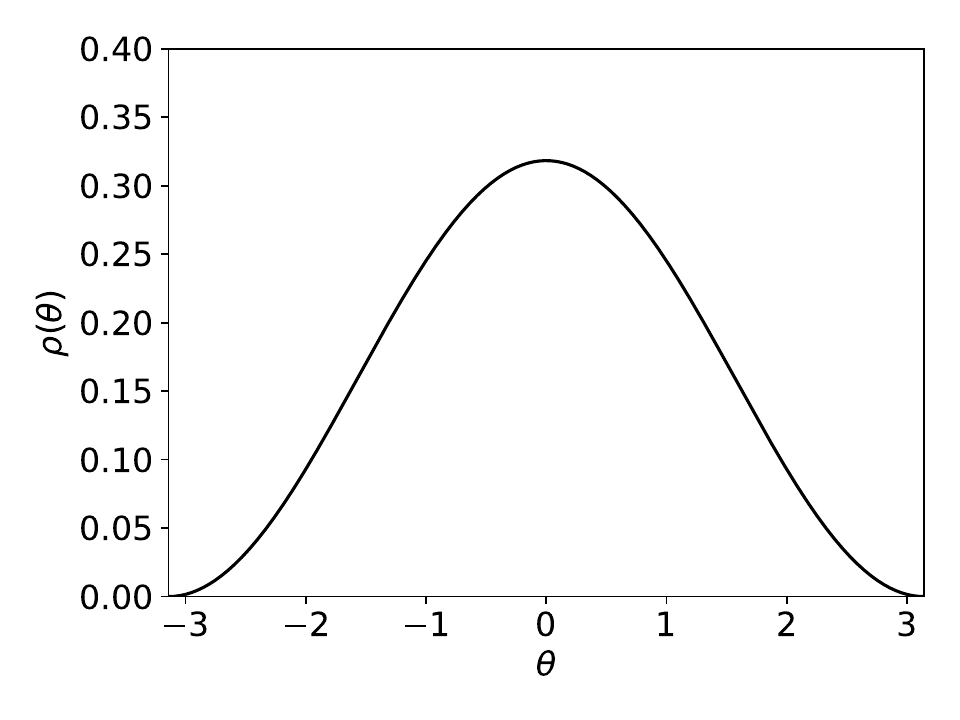}
     \caption{The critical distribution marking the Gross-Witten-Wadia transition between gapped and ungapped phases.
     At zero chemical potential, this marks the boundary between partial and complete deconfinement. At finite chemical potential, this critical point might instead correspond to the condensation of particles with non-trivial gauge orbits, such as baryons.}
     \label{fig:gww-transition}
\end{figure}

The critical point between the gapped and ungapped phases is known as the Gross-Witten-Wadia (GWW) point, depicted in Fig.~\ref{fig:gww-transition}. At zero chemical potential, the corresponding GWW transition is therefore identified with the transition between complete and partial deconfinement. 

\subsubsection{GWW transition and nontrivial condensates} \label{sec:nontrivial-gauge-condensates-GWW}

Historically, the GWW transition was regarded as the definitive signature of the partially deconfined phase. However, it was recently noticed that the GWW transition can also be induced by the condensation of objects with nontrivial gauge orbits in the extended Hilbert space \cite{Hanada:2025rca}. Then, the GWW transition reflects the condensation of these objects instead of the onset of partial deconfinement. A prominent example is provided by baryons in QCD, which can be induced to condense by the presence of a chemical potential. As a result, the GWW transition can occur while still entirely within the partially deconfined phase, and can no longer be regarded as a conclusive indicator of partial deconfinement. The GWW transition in the BPS phase diagram cannot therefore be used by itself to identify the partially deconfined phase. As we will see, this realisation resolves a paradox about the location of the partially deconfined phase in the BPS phase diagram. 

\subsubsection{Gauge redundancy and Bose-Einstein interference} \label{sec:pdec-bec}
When a quantum state exhibits invariance under a nontrivial subgroup of the gauge group, its contribution to physical observables can be statistically enhanced. This is analogous to the enhancement of configurations of identical bosons that sit in the same state, as found in the ground state of Bose-Einstein condensates. There, the invariance of such bosons under permutation symmetry can be regarded as a gauge redundancy. This effective enhancement effect drives confinement at weak coupling.

To see this quickly, note that states in the physical (i.e. gauge-invariant) Hilbert space can be described by a projection over the gauge group on the extended (i.e. gauge-dependent) Hilbert space $\mathcal H_{\rm ext}$. In this way, the contribution of a state $\ket{\phi}$ to the partition functions is,
\begin{align}
    e^{-E_\phi/T} \frac{1}{\rm{vol}~G} \int_G dg \Braket{\phi|\hat g|\phi},
\end{align}
where $E_\phi$ is the energy of the state. Here, the action of $\hat g$ describes the holonomy over the Euclidean time direction \footnote{It is sufficient to integrate a single copy of the gauge group \cite{Hanada:2020uvt}.} If, for example, the state $\phi$ is invariant under a subset $SU(N-M)$, then this integral picks up an enhancement from the volume of this subgroup, here $e^{N^2-M^2}$. If the state has no such invariance, then the contribution of this state to the partition function does not experience such an enhancement effect. This is explained in more detail in \cite{Hanada:2020uvt}. See also Sec.~2.1 of \cite{Hanada:2025rca}.

The important lesson for our later matrix model analysis is that this enhancement factor is captured by the Jacobian (the Vandermonde term), which measures the gauge orbit volume associated with a given eigenvalue configuration. This is the sole term responsible for eigenvalue repulsion, and therefore drives both confinement and, as we will see, instanton condensation. Acknowledging this will help with the physical interpretation of our results for the BPS phase diagram later.

\subsection{Instantons} \label{sec:pdec-instantons}
Since the GWW transition does not provide a universal diagnostic of partial deconfinement, it is natural to seek alternative characterisations. One such characterisation is provided by instanton condensation in matrix models. In general, the partition function of any quantum field theory is expected to admit a transseries of the schematic form,
\begin{equation} \label{eq:Z-instanton-expansion}
    Z = Z_0 + \sum_i Z_i e^{-\Delta S_i}.
\end{equation}
The exponentially-suppressed contributions proportional to $e^{-\Delta S_i}$ are not captured by perturbation theory. These contributions could originate from topological objects such as monopoles and instantons, or from renormalons. The word `instanton' in this context is used to cover all such exponential contributions. 

In matrix model reductions of Yang-Mills theories at zero chemical potential, the GWW transition coincides with the behaviour,
\begin{equation} \label{eq:delta-S-0}
    \Delta S_i \rightarrow 0
\end{equation}
for some $i$, such that the exponential contribution is no longer suppressed and cannot be neglected in the region of the transition \cite{Neuberger:1980qh,Neuberger:1980as,Marino:2008ya,Buividovich:2015oju}. This is described as \emph{instanton condensation}. It is often argued that instanton condensation drives or triggers the transition \cite{Neuberger:1980as,Marino:2008ya}. It seems plausible that these instanton contributions represent the remnants of extended topological objects, such as genuine instantons, under dimensional reduction of the original Yang-Mills theory. This idea was tested in \cite{Hanada:2023krw, Hanada:2023rlk} for WHOT-QCD, a lattice theory that closely mimics QCD, but with heavier up and down quarks. It was confirmed that a nontrivial distribution of topological charge condenses as one moves from the deconfined to the partially deconfined phase, indeed signalling the condensation of actual topological objects. 

Instantons have separately been implicated in mechanisms of chiral symmetry breaking. Another longstanding prediction of the partially deconfined phase is that chiral symmetry breaking occurs at the transition point. This can be argued, for example, using the famous 't Hooft anomaly between centre and chiral symmetry.
Instanton condensation might therefore play a very interesting and important physical role in partial deconfinement. Furthermore, the behaviour \eqref{eq:delta-S-0} is associated with Stokes and anti-Stokes phenomena, which can describe first and higher-order phase transitions \cite{Fujimori:2021oqg, Dunne:2012ae, Marino:2007te}. 
Instanton condensation as a feature of the partially deconfined phase will play a central role in our analysis of the superconformal index and search for the dual of the partially deconfined phase.

\subsection{Thermodynamics}
\begin{figure}  
    \centering
    \begin{subfigure}[b]{0.48\textwidth}
    \centering
    \includegraphics[width=\linewidth]{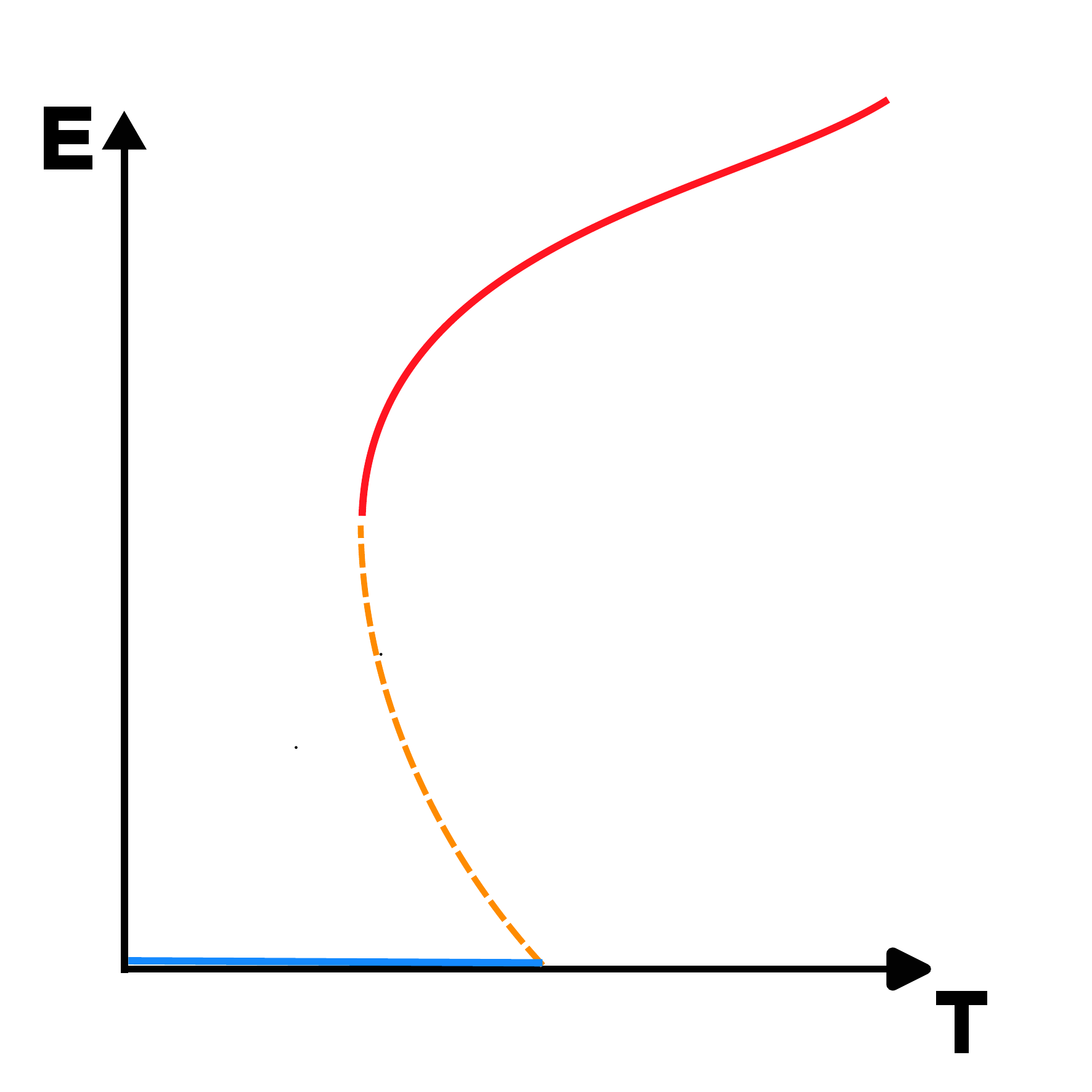}
    \caption{}
    \label{fig:pdec-energy-temperature}
    \end{subfigure}
    \hfill
    \begin{subfigure}[b]{0.48\textwidth}
    \centering
    \includegraphics[width=\linewidth]{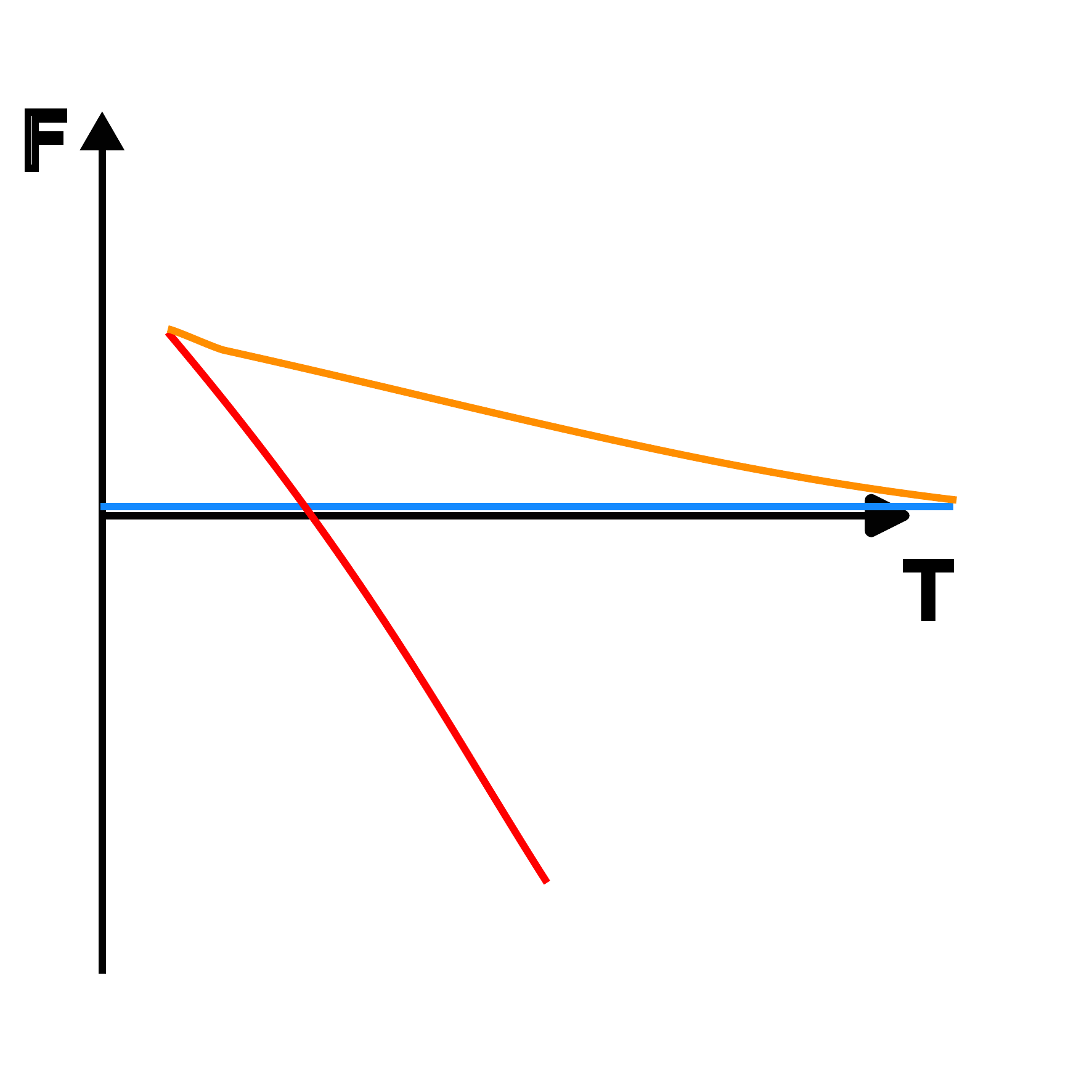}
    \caption{}
    \label{fig:pdec-freeEnergy-temperature}
    \end{subfigure}
    \caption{The confined (blue), deconfined (red), and partially deconfined (orange) saddles in the situation where the partially deconfined phase has negative heat capacity. This shows (\subref{fig:pdec-energy-temperature}) energy and (\subref{fig:pdec-freeEnergy-temperature}) free energy, respectively, against temperature. This depicts a first order phase transition in the canonical ensemble, from the confined to deconfined phases. In the microcanonical saddle, where we fix energy, the partially deconfined phase becomes stable. The partially deconfined saddle necessarily connects the confined and deconfined saddles.}
    \label{fig:pdec-thermodynamics-sketch}
\end{figure}

The confined and deconfined saddles are continuously connected through the partially deconfined saddle, as demonstrated in Fig.~\ref{fig:pdec-thermodynamics-sketch}.
As a greater volume of the gauge group leaves the confined phase and enters the deconfined phase, macroscopic variables, such as energy, likewise interpolate between the fully confined and fully deconfined values.
In many situations, the partially deconfined phase has negative heat capacity. A negative heat capacity is rare in field theories, but the partially deconfined phase explains its origin naturally. As the energy increases, more degrees of freedom become active, since more of the colour space becomes deconfined. The value $C_V = T \left(\frac{\partial^2 E}{\partial S^2}\right)^{-1}$ then often becomes negative, depending on the details of the theory under consideration \cite{Hanada:2018zxn}.

It is also possible for the saddle to have positive specific heat, typically depending on the matter content \cite{Aharony:2003sx, Aharony:2005bq}. In the interpretation of the QGP crossover as partial deconfinement, it is thought that the quarks are responsible for shifting the saddle so that it has positive specific heat and is stable. This captures the observed features of the crossover.

\subsection{Proposed holographic duals} \label{sec:pdec-ggd}

The partially deconfined phase was discovered initially through an attempt to find a field theory dual explanation for certain small black hole configurations, particularly those with negative specific heat. The large black hole in $AdS_5 \times S^5$ has long been known to be dual to the standard deconfined phase \cite{Witten:1998qj}. The small Schwarzschild black hole, meanwhile, is distinguished from the large black hole by its negative heat capacity. The two saddles exhibit the same relationship between complete and partial deconfinement as proposed in Fig.~\ref{fig:pdec-thermodynamics-sketch}, and meet at a cusp as in Fig.~\ref{fig:pdec-energy-temperature}. It was therefore conjectured in \cite{Hanada:2016pwv} that the small black hole is dual to the partially deconfined phase.

In various systems analysed at weak coupling, the GWW transition coincides with the change in sign of the specific heat, supporting this proposal \cite{Hanada:2018zxn,Hanada:2019kue,Aharony:2003sx,Sundborg:1999ue,Schnitzer:2004qt}. 
Moreover, computations on the BFSS and BMN matrix models at strong coupling have been shown to obey this pattern, with the entire negative heat capacity saddle identified with the partially deconfined phase \cite{Bergner:2021goh}.

In the case of thermal black holes in $AdS_5 \times S^5$, additional structure is present. It is known that sufficiently small black holes undergo a Gregory-Laflamme instability, leading to black hole configurations that are localised on the compact $S^5$. Under the reasoning described above, both the small Schwarzschild black holes and the localised black holes are regarded as partially deconfined phases \cite{Hanada:2018zxn,Hanada:2019kue}. However, the evidence for identification with partial deconfinement is strongest in the regime of localised black holes, where the correct thermodynamic scaling laws and a direct correspondence with a reduced number of active colour degrees of freedom can be established \cite{Berenstein:2018lrm, Hanada:2016pwv}. The extension of this identification to the entire small–black–hole saddle is comparatively indirect.

Furthermore, the identification of the entire negative heat capacity saddle presents a significant puzzle. From the bulk geometric viewpoint, the transition from the large to the small black hole seems to be smooth, with all thermodynamic potentials continuous at the cusp and no other features usually associated with a sharp phase transition. By contrast, the field theory description of the transition to partial deconfinement is expected to involve the condensation of instantons and a genuine phase transition.

Confusion therefore remains as to which parts of the black hole phase diagram correspond to the partially deconfined phase and how the transition is represented in the bulk geometry. Resolving this question in the case of BPS black holes is a primary motivation for the present work.

\section{Superconformal index matrix model} \label{sec:index}
We will present here the central tool of our analysis, the superconformal index. This generalises the Witten index \cite{Witten:1982df}, employing the state-operator map to make similar arguments applicable to quantum field theories possessing superconformal symmetry \cite{Kinney_2007,R_melsberger_2006}. The index counts BPS states, which are those satisfying
\begin{equation}
    \Q_\alpha \ket{\psi} = 0,
\end{equation}
up to grading by fermion number. We will focus on the case of $\mathcal N=4$ SYM, which we will regard as an $\N=1$ theory in order to count 1/16-BPS states, treating the additional supersymmetries as ordinary non-supersymmetric symmetries. We can also insert chemical potentials for charges that commute with $\Q$. In $\N=4$ SYM, the commuting charges consist of the two angular momenta $J_k$ and the three Cartan charges $R_i$ in the $SU(4)$ R-symmetry. In the gravitational dual theory, the $R$ charges correspond to the three Cartan subalgebra elements of the $SO(6) \cong SU(4) $ rotations of the compact $S^5$. Our expression for the superconformal index of $\mathcal N=4$ SYM on $S^3 \times S^1$ can then be written,
\begin{equation}
    Z =\Tr_{ \mathcal H_{S^3} } \left[ {(-1)^F e^{-\beta \{ \Q, \Q^\dagger \} } e^{-\sum_{i=1}^{3}\Delta_i Q_i} e^{-\sum_{k=1}^{2}\omega_k J_k}} \right],
\end{equation}
with a constraint on the chemical potentials, $\Delta_1 + \Delta_2 + \Delta_3 - \omega_1 - \omega_2 = 2\pi i k$, where $k$ is a choice of integer \cite{Kinney_2007,choi2024yang}. We must take periodic boundary conditions for the fermions around the $S^1$, with circumference $\beta$, to preserve supersymmetry. This index is invariant under $\beta$ since, by the standard arguments, only states satisfying $\{ \Q, \Q^\dagger \}=0$ contribute. Moreover, it is invariant under RG-flow and changes of coupling strength. We can therefore calculate the index accurately at strong coupling, the domain relevant to the gravitational dual, by computing the index in the weak coupling limit, using weak coupling field theory techniques.

Throughout this work, we shall restrict to a simpler case, setting $e^{-\Delta_1}=e^{-\Delta_2}=e^{-\Delta_3}=x^2$ and $e^{-\omega_1}=e^{-\omega_2}=x^3$. The variable
$x$ is now the only independent variable controlling the fugacities, and can be thought of as measuring a `supersymmetric temperature' $\tau$ via $x= e^{-1/\tau}$. To preserve supersymmetry, we cannot introduce a true temperature, as antiperiodic boundary conditions for fermions would break supersymmetry. Note that the chemical potentials are periodic under shifts by $2 \pi i$, and the value of the integer $k$ in the above constraint is now meaningful only modulo 3. Our choice of definition makes $x$ single-valued in the complex plane, with all three branches of $k$ accessible by values of $x$ in the complex plane.

Our approach will be to compute the index via the path integral formalism in the weak coupling limit, following \cite{Aharony:2003sx,choi2024yang}. While other modes contribute at one-loop level, the gauge field $A_0$ pointing in the temperature direction contains a zero mode that must be handled exactly. We should therefore compute an integral over the holonomy matrix, $U$, over the temperature direction, which is the Polyakov line.

To carry out this integral in practice, we will represent the holonomy by its eigenvalues, $e^{\alpha_i}$. The measure must then combine with a Jacobian, which is known as the Vandermonde determinant,
\begin{equation}
    \int [dU] \rightarrow \frac{1}{N!} \prod_{i=1}^N \int \frac{d \alpha_i}{2\pi} \prod_{1\leq j<k \leq N} \lvert e^{i\alpha_j} - e^{i\alpha_k} \rvert^2.
\end{equation}
This represents the volume element of the gauge orbit for a given choice of eigenvalues. It acts as a repulsion term on the eigenvalues, encouraging confinement. As explained in Sec.~\ref{sec:pdec-bec}, it makes sense that the term measuring gauge redundancy should be the crucial factor driving confinement, including the confining properties of partial confinement. Remembering the physical origin of this contribution will be useful for us when interpreting our results later.

Later, we will need to find \emph{complex} saddles of the index. Therefore, it will be more convenient to write the action in terms of complex variables $z_k=e^{i\alpha_k}$. Furthermore, since we are in the large $N$ limit, it will be useful to write this integral in terms of the eigenvalue density, analogous to eq.~\eqref{eq:rho-density-defn}, 
\begin{equation}
    \rho(\zeta) = \frac{1}{N} \sum_{i=1}^N \delta(\zeta-z_j).
\end{equation}
The notion of the continuum limit can be made more precise with resolvents, which show that we can take $z$ to be continuous and turn discrete sums into weighted integrals,
\begin{equation}
    \rho(\zeta) \Delta \zeta \rightarrow \rho(z) dz.
\end{equation}
As a density distribution, $\rho(z)$ is required to be real and nonnegative, with the normalisation constraint,
\begin{equation} \label{eq:rho-normalisation}
    \int_{\bigcup C_k} \rho(z) = 1,
\end{equation}
where $C_k$ labels all the disjoint contours in the complex plane on which $\rho(z)$ is supported. For real saddles, the $z_i$ and hence contours $C_k$ are taken to sit on the unit circle. When we deal with complex saddles, however, we must allow them to sit more generally on the complex plane. The way in which these contours will be selected will be described later.

Combining everything, we obtain
\begin{align}
    Z(x)=\prod_{i=1}^N \int_C dz_i \exp(-N^2 S_N[\rho[\{z\}]])
\end{align}
with action\footnote{the term $\log(z-z')^2$ should be understood as the sum of limits taken on each side, $\log(z_+-z') + \log(z_--z')$, for $z,z' \in C_k$, with appropriate branch cuts for the logarithm},
\begin{equation} \label{eq:action}
       S_N[\rho]
    =
    \frac{1}{\lambda} \int_{\cup C_k} \lvert dz \rvert W(z) \rho(z)  - \frac{1}{2} \int_{\cup C_k} \lvert dz \rvert  \int_{\cup C_k} \lvert dz' \rvert
    \log(z-z')^2 \rho(z) \rho(z')    
\end{equation}
and
\begin{equation}
    W(z) = \log z - \sum_{m=1}^\infty \frac{a_n(x) \rho_n}{n} (z^n + z^{-n}),
\end{equation}
where
\begin{equation} \label{eq:def-an-rho}
    \rho_n=\int_\gamma dz~ z^n \rho(z), \qquad a_n(x) = 1 - \frac{(1-x^{2n})^3}{(1-x^{3n})^2}.
\end{equation}
We have separated $W(z)$, which has the appearance of an external potential on a single eigenvalue, from the second term, which has the form of an interaction between eigenvalues. The interaction term is repulsive and comes entirely from the Vandermonde determinant. The `external potential' depends on the entire distribution of eigenvalues through $\rho_n$, and so finding solutions is not as trivial as this form might suggest. Note that the dependence on the fugacity $x$ enters only through $a_n(x)$.

\subsection{Saddle point solutions}

We can solve the equations of motion with the constraint \eqref{eq:rho-normalisation} to obtain
\begin{equation} \label{eq:v-eom}
    V(z) = \frac{\delta S_N[\rho]}{\delta \rho(z)} = \frac{1}{\lambda}W(z)-\int_{\cup C_k} \log(z-z')^2 \rho(z') |dz'| = \mu,
\end{equation}
for a constant Lagrange multiplier $\mu$.
This has a form similar to electrostatics in 2D. From this perspective, the action $S_N[\rho]$ can be interpreted as a complex extension of electrostatic energy, $V(z)$ as a complex extension of the electric potential, and $\rho(z)$ as the charge density. Then, eq.~\eqref{eq:v-eom} is the statement that the electric potential is constant along the contour. We can define an electric field over the complex plane by,
\begin{equation}
    y(z) = -\overline{V'(z)},
\end{equation}
where the bar denotes complex conjugation.
This can give immediate intuition about some relations that are presented more rigorously in e.g. \cite{Alvarez:2016rmo}. Contours with `charge' occupation correspond to discontinuities in the electric field that they source. The electric field on each side is equal and opposite,
\begin{align}
    y(z_+) + y(z_-)=0, \qquad z \in C_k,
\end{align}
the forces on each side balancing to achieve equilibrium on the charges. The endpoints of the contours, where the charge density reaches zero, must be situated where the electric field, or $y(z)$, vanishes. We will call these zeros \emph{stationary points} of $y(z)$. When we have a double root of $y(z)$ which does not mark the endpoint of a contour, we can also call them \emph{pinching points}. The potential along each contour must be constant. 
When defined more rigorously, $y(z)$ is known as the spectral curve. Continuing with the analogy with electrostatics, $y(z)$ is multivalued or ill-defined on the contour where the charge density $\rho(z)$ is defined, and can relate it to the discontinuity with the canonical relation,
\begin{align} \label{eq:y-rho-relation}
     \rho(z) = \frac{1}{4 \pi i} \mathrm{Disc}~y(z), \qquad z\in \cup_k C_k
\end{align}
The potential difference between a contour and any point $z$ in the complex plane can be measured by
\begin{align} \label{eq:Veff-integral}
    V_{\rm eff}(z) = - \re~ \int_a^z y(z')\, dz',
\end{align}
where $a$ can be any point of the contour, usually taken to be one of the endpoints. A particularly significant situation to consider is the potential between a cut and a pinching point. 
Evaluating \eqref{eq:Veff-integral} with $z=b$ as this pinching point then shows whether it is favourable for charge density to move from one point to the other, with $V_{\rm eff}<0$ signalling an instability of the saddle. Moving the filling fraction $\delta \nu$ (that is, the quantity of `charge', where total charge is $\int_{\cup C_k} \rho=1$) from the contour to this pinching point, the real part of the action will change by approximately,
\begin{equation}
    \Delta S \approx V_{\rm eff}(b) \delta \nu,
\end{equation}
and contribute a term proportional to $e^{-N^2\Delta S}$ to the partition function. When $\Delta S > 0$, this term is exponentially suppressed against the leading term, as expected for an instanton. Remembering the discrete origin of the eigenvalue density, we can consider a finite number of eigenvalues tunnelling to the region around the pinching point, opening up the new cut. Such configurations will contribute nonperturbative terms to the partition function transseries \eqref{eq:Z-instanton-expansion}, with the index $i$ labelling the number of tunnelling eigenvalues. We conclude that the expression \eqref{eq:Veff-integral} can be used to obtain the instanton contributions to the matrix model partition function that describes the index in the sense discussed in Sec.~\ref{sec:pdec-instantons}. We hence identify the instanton action per unit filling fraction, $A_b$, corresponding to eigenvalues tunnelling to a pinching point at $b$ as,
\begin{equation}
    A_b = V_{\rm eff}(b).
\end{equation}
In the literature, this is often known simply as the instanton action \cite{Marino:2008vx, Marino:2008ya}. Since the first tunnelling instanton corresponds to a filling fraction $\delta\nu=1/N$, the associated contribution to the partition function scales as $ e^{-N A_b}$.

\subsubsection{Single-cut solutions} \label{sec:singel-cut-solution-definition}
A set of solutions of the saddle point equations for this effective action have been obtained in previous work \cite{choi2024yang}. These assume a single cut with two distinct endpoints and symmetry under $\alpha \rightarrow -\alpha$, i.e. $z \rightarrow 1/z$ on the complex plane. The focus of our current paper will be to apply the above effective action to analyse the stability of these known solutions against instanton condensation, which determines when alternative saddles become dominant. For completeness, we now summarise how these solutions are constructed.

We define the vector of moments of the density distribution up to a truncation, $p$, by  $\vec \rho=(\rho_1,\rho_2,\cdots,\rho_p)$, taking the single contour in our ansatz as the domain for the integral in eq.~\eqref{eq:Veff-integral}. It was shown in \cite{choi2024yang, Aharony:2003sx} that the saddle equation can be solved for a given truncation via the expression,
\begin{equation}\label{rho-formal-sol}
\vec{\rho}=M_{p \times p}^{-1}\vec{e}_1\ ,
\end{equation}
where $e_1$ is a unit column vector of length $p$ given by $(1,0,0,\cdots)$ and $M_{p \times p}$ is a $p \times p$ matrix constructed as follows. We first define,
\begin{eqnarray}\label{R}  R_{ml}&=&a_l\sum_{k=1}^l\left[B^{m+k-\frac{1}{2}}(s^2)+
  B^{\left|m-k+\frac{1}{2}\right|}(s^2)\right]P_{l-k}(1-2s^2)\nonumber\\
  A_m&=&a_m\left[P_{m-1}(1-2s^2)-P_m(1-2s^2)\right]\ .
\end{eqnarray}
where $s^2\equiv\sin^2\frac{\alpha_0}{2} $
and
\begin{eqnarray}
  B^{n-\frac{1}{2}}(s^2)&=&\frac{1}{\pi}\int_{-\alpha_0}^{\alpha_0}
  d\theta\sqrt{s^2-\sin^2\frac{\theta}{2}}~\cos\left[
  \textstyle{\left(n-\frac{1}{2}\right)}\theta\right]\nonumber.
\end{eqnarray}
We can determine $s^2$ in terms of the $a_n$ variables by solving,
\begin{align}
    \det(R-\boldsymbol{1})=0.
\end{align}
Then we construct,
\begin{equation}
M_{p \times p} =
\begin{bmatrix}
A_1 & A_2 & \cdots & A_p \\
(1-R)_{21} & (1-R)_{22} & \cdots & (1-R)_{2p} \\
\vdots & \vdots & \ddots & \vdots \\
(1-R)_{p1} & (1-R)_{p2} & \cdots & (1-R)_{pp}
\end{bmatrix},
\end{equation}
i.e. we replace the first row of the matrix $({\bf 1}-R)$ by the vector $\vec{A}$. Thus, we obtain the moments $\vec \rho$. Finally, substituting the resulting distribution $\tilde \rho(z)$ into \eqref{eq:action}, we obtain the saddle point free energy,
\begin{equation}
    Q =S_N[\tilde \rho].
\end{equation}

However, even after solving the above equations, the existence of a fully-consistent saddle solution is not guaranteed. The requirement of real nonnegative $\rho$ along the cuts demands that the function,
\begin{align}
    G(z) = \int_{a}^z  y(z) dz,
\end{align}
taken along the cuts must satisfy,
\begin{equation}
    \lim_{z \rightarrow z_\gamma} {\rm Re}~G(z_\gamma) =0, \qquad z_\gamma\in \cup C_k
\end{equation}
for all points $z_\gamma$ on the cut~\cite{David:1990sk}.
In this way, we can define the contours $C_k$ and their embedding in the complex plane. Stationary points satisfying $y(z)=0$ mark the ends of the contours. A contour can split as we change $x$ if a stationary point passes through it \cite{choi2024yang}.
If this occurs in the context of our single-cut ansatz, the assumption of a single cut is violated, and the above solution is no longer valid. This condition is very nontrivial, and in general must be checked numerically.

Moreover, determining whether complex saddles contribute to a contour integral of the path integral is not obvious. A thorough assessment would involve, for example, Lefschetz thimble analysis. This would be very complicated in the current setup, where we are actually integrating over a large $N$ number of variables, and is usually avoided. Instead, physical reasoning is often used to justify accepting or rejecting a complex saddle as contributing. This is the approach we shall mostly follow in this work. However, we will also make use of some direct numerical evaluation to confirm that certain complex saddles most crucial to our analysis indeed contribute to the integral. 

\subsection{BPS black holes} \label{sec:BPS-black-holes}
To obtain the black hole phase diagram from the superconformal index, it was prescribed in \cite{Cabo_Bizet_2019,choi2024largeadsblackholes} that we should find the charge, $q = x \frac{\partial{Q}}{x}$, and require this charge to be real, $\mathrm{Im}~q = 0$. This is the condition to find the physical black holes saddles. Then, we can perform the Legendre transform to obtain the entropy,
\begin{equation}
    S = Q - q \log x.
\end{equation}
The saddle with the greatest real part of those that actually contribute to the contour integral is taken to be dominant. 

\begin{figure} 
    \centering
    \begin{subfigure}[b]{\textwidth}
    \centering
    \includegraphics[width=0.5\linewidth]{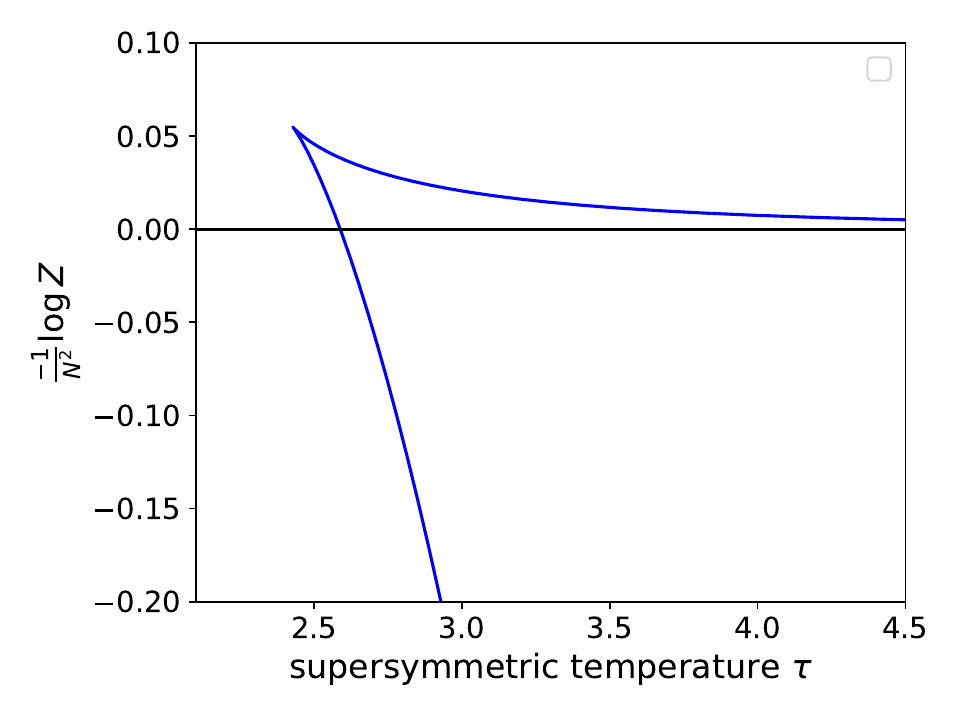}
    \end{subfigure}
    \caption{Phase diagram of the BPS black hole free energy against supersymmetric temperature $\tau$, made using the results of \cite{Ezroura_2022}. The cusp around $\tau \approx 2.4$ marks the point at which the large and small black hole branches meet.} \label{fig:BPS-black-hole-free-energy-Ezoura}
\end{figure}

Following this prescription with the solution detailed in the previous section shows that the single-cut symmetric solutions describe the Gutowski-Reall $1/16$-BPS black hole solutions well \cite{Gutowski:2004ez, Gutowski:2004yv}. Computations using the index and direct supergravity calculations have established that the BPS black hole phase diagram closely follows what is expected for the thermal AdS Schwarzschild black hole, with the supersymmetric temperature $\tau$ behaving analogously to temperature. In particular, there is again a large black hole solution, a small black hole solution characterised by negative heat capacity, and a thermal gas \cite{Ezroura_2022,choi2024yang,Copetti:2020dil} sitting at zero free energy at $O(N^2)$. The small black hole connects to the large black hole at a cusp, and smoothly connects to the thermal gas in the infinite $\tau$ limit. A Hawking-Page transition occurs in the grand canonical ensemble between the large black hole and the AdS gas, and this describes the deconfinement transition. The small black hole is never dominant in the grand canonical ensemble. The gravity solution provided in \cite{Ezroura_2022} for the equal-charge BPS black hole is reproduced in Fig.~\ref{fig:BPS-black-hole-free-energy-Ezoura}.

A notable difference between the known BPS black hole phase diagram and the phase diagram for thermal black holes is the presence of a Gregory-Laflamme transition in the thermal case. This corresponds to black holes localised on the compact $S^5$, and becomes dominant in the microcanonical ensemble at sufficiently low radius or mass. In \cite{Bhattacharyya:2010yg}, it was argued that BPS black holes are stable against such Gregory-Laflamme transitions, based on reasoning that assumes a small 10d Schwarzschild black hole as the endpoint of the transition at very small charge.
The results we present later will encourage us to tentatively reconsider this argument and its assumptions.

We will be most interested in the small black hole branch. Therefore, we will work in the microcanonical ensemble, in which the charge $q$ is fixed, and the saddle with greatest entropy dominates.

\section{Matrix model truncations} \label{sec:truncations}

To evaluate single-cut solutions of the superconformal index according to the method described in Sec.~\ref{sec:singel-cut-solution-definition}, we must pick a truncation $p$. Higher values of $p$ lead to greater accuracy of the approximation, but also increase computational complexity. The number $p$ describes the number of modes to include in the spectral curve, and therefore must take the general form,
\begin{align} \label{eq:y-general}
    y(z)^2 = g_p^2 \frac{(z-a)(z-a^{-1})(z+1)^2 \prod_{i=1}^{p-1} (z-d_i)^2(z-d_i^{-1})^2}{z^{2p+2}},
\end{align}
where $g_p$ is the coupling. Carefully working through the construction in Sec.~\ref{sec:singel-cut-solution-definition} shows that the coupling should be given by $g_p = a_p \rho_p$.

In this section, we will examine the first three truncations in more detail. The first truncation, $p=1$, already contains all the essential features of the instanton instability that we will later demonstrate for the BPS black hole, and it is therefore worth understanding in detail. Moreover, the $p=1$ model also happens to be the effective description for Yang-Mills theories on $S^3 \times S^1$. Lessons drawn from this relatively simple example will assist in the physical interpretation of our results for the black hole and its field theory dual.

Higher truncations exhibit additional complexities, such as new pinching points and instantons. These will also be discussed.

\subsection{The $p=1$ truncation (Gross-Witten-Wadia model)} \label{sec:matrix-model-detail-gww}
\subsubsection{Real coupling: gapped and ungapped phases}
The $p=1$ truncation is the original Gross-Witten-Wadia model \cite{Gross:1980he, Wadia:2012fr}. Its phase structure is well-understood. The single-cut gapped $y(z)$ simplifies to,
\begin{equation}
    y(z)^2 = g^2\frac{(z+1)^2(z-a)(z-a^{-1})}{z^{4}},
\end{equation}
where we set $g \equiv g_1$ for clarity. Solving the saddle equations shows that, for real coupling and $g>1$, the endpoint depends on the coupling as,
\begin{equation}
    a = g^{-1}+\sqrt{g^{-2}-1}.
\end{equation}
At real coupling, this describes a contour that covers an arc on the unit circle and is therefore a gapped distribution according to Fig.~\ref{fig:gww-phases}. The free energy of the configuration can be calculated by inserting this solution into \eqref{eq:action} (using \eqref{eq:y-rho-relation}) and performing the integration. This yields the well-known formula,
\begin{equation} \label{eq:free-energy-G}
    F_G(g) = g - \frac{1}{2} \log g - \frac{3}{4}.
\end{equation}
There is a pinching point in the complex plane at $z=-1$, evidenced by the root of $y(z)$ here. By integrating \eqref{eq:Veff-integral} between the endpoint of the cut $a$ and this critical point, we obtain the action for the eigenvalue tunnelling to this point,
\begin{equation} \label{eq:instanton-G}
    A_G(g) = \int_a ^{-1} y(z) dz = 2 \sqrt{g(g-1)} - \cosh^{-1}(2g-1).
\end{equation}
This is the instanton action for the gapped phase.
Again, at real coupling, this solution is valid at $g>1$.

For $g<1$, the solution is instead an ungapped distribution, meaning $\rho(\theta)$ is nonzero over the entire unit circle. Since the cut no longer has two distinct endpoints, the general form \eqref{eq:y-general} no longer applies. Instead, $y(z)$ now takes the form,
\begin{equation} \label{eq:y-p=1-ungapped}
    y(z)^2 = g^2\frac{(z-a)^2(z-a^{-1})^2}{z^{4}}
\end{equation}
where $a$ can again be written as a function of $g$, specifically,
\begin{equation}
    a = 1-2g^{-1}+2\sqrt{\frac{1-g}{g^2}}.
\end{equation}
The free energy can be calculated by the same methods to give,
\begin{equation} \label{eq:free-energy-UG}
    F_{UG}(g) = \frac{g^2}{4}.
\end{equation}
Now, the values of $a$ and $1/a$ lie on the imaginary axis at $z=-1$, with one point inside the unit circle and the other outside. As shown by \cite{Buividovich:2015oju,Alvarez:2016rmo}, the concept of eigenvalues tunnelling to these points to open new cuts, and the corresponding actions being interpretable as instanton actions, continues to be valid. The Coulomb gas method of obtaining the instanton action by integrating $y(z)$ between the cut and $a$ or $1/a$ again agrees with the independent orthogonal polynomial method of \cite{Marino:2008ya}.
However, in order for equilibrium to be maintained, we recognise that the instantons must tunnel in pairs, effectively doubling the instanton action.

Before continuing, we must note an important subtlety when using the Coulomb gas approach to compute instanton actions. In essence, the naive computation for one of the pair of instantons can sometimes take the wrong sign. Wrong-sign instantons are known as \emph{ghost instantons}. It was shown in \cite{Marino:2022rpz} that these can be understood as eigenvalues tunnelling to an non-physical sheets of the spectral curve in the complex plane. Intantons tunneling to the pinching point inside the unit circle in the ungapped distribution is one example of this. Recognising this, we finally obtain,
\begin{equation} \label{eq:instanton-UG}
    A_{UG} = \int_{-1} ^{a} y(z) dz - \int_{-1} ^{a^{-1}} y(z) dz = 2\cosh^{-1}\left( \frac{1}{g} \right) - 2\sqrt{1-g^2},
\end{equation}
for the instanton in the ungapped phase.

At $g=1$, the gapped and ungapped phases meet smoothly, giving the famous third-order GWW transition. Away from $g=1$, the instanton actions are always positive, leading to exponential suppression of the instantons. However, as we approach $g=1$, both expressions for the instantons begin to vanish, and their contribution can no longer be neglected, as in \eqref{eq:Z-instanton-expansion} and \eqref{eq:delta-S-0}. It is in this sense that we can say that instanton condensation drives the GWW transition.

If eigenvalues tunnel to the pinching points, new cuts open up, and we obtain multi-cut distributions \cite{Buividovich:2015oju,Alvarez:2016rmo,Marino:2007te}. At real coupling, away from the critical point $g=1$, the instanton actions are always positive and multi-cut distributions are always suppressed. Therefore, we conclude that only single cut solutions dominate thermodynamically at real coupling.

\subsubsection{Complex coupling and multi-cut saddles}

Computing complex saddles of BPS black holes will require extending the truncated matrix models to complex values of the couplings. The full generalisation of these models to complex coupling is not yet fully understood. However, some important features can be readily deduced.

We expect that the expressions provided in this section, such as the free energy and the locations of the cut endpoints, can be analytically continued to complex coupling if we can continuously deform the original contour into the complex plane
such that $\rho(z)$ is everywhere real and nonnegative on the new contour.
This can be checked numerically for any truncation, and in the case of $p=1$, an explicit reparametrisation of the contour for the analytically-continued ungapped phase is given in \cite{Copetti:2020dil}.
Therefore, we can be confident that the expressions 
for the free energy and instanton actions given above will continue to apply for complex values of $g$.

\begin{figure}
    \centering
    \includegraphics[width=0.5\linewidth, keepaspectratio]{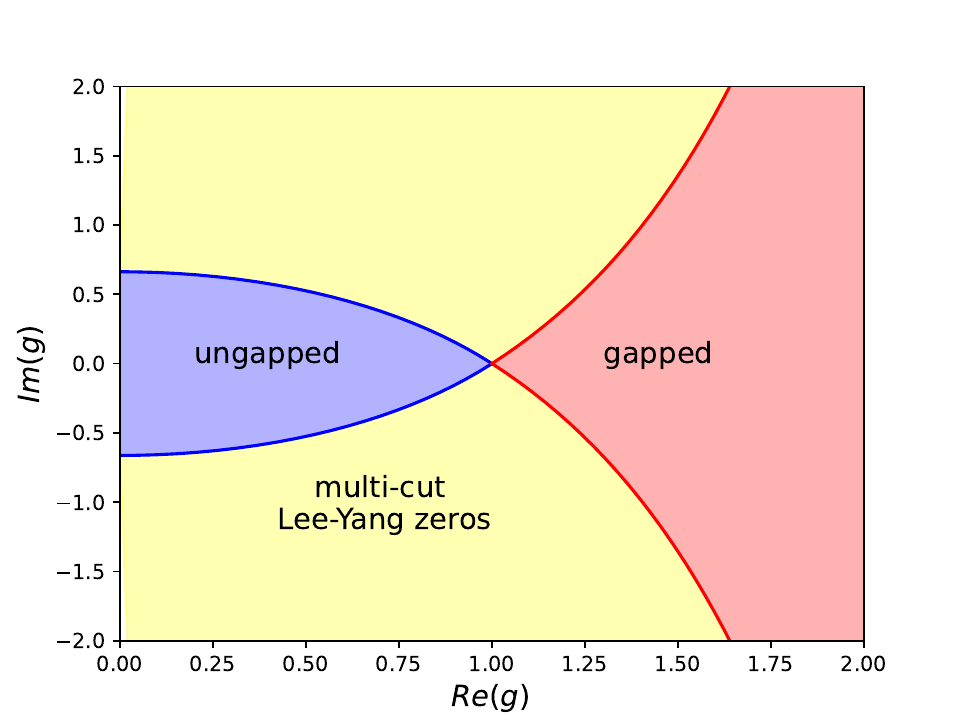}
    \caption{The phase diagram of the Gross-Witten-Wadia model at complex coupling. The regions where the single-cut gapped and ungapped solutions dominate are shown in red and blue, respectively. These are bounded by instanton condensation curves. This indicates where dominance by multi-cut saddles is expected to begin, shown in yellow. The gapped and ungapped instanton condensation curves meet on the real axis at the GWW point but separate in the complex plane. The Lee-Yang zeros also pinch the real axis at this point, as first observed by \cite{Kolbig:1981qz}. We expect the Lee-Yang zeros to fill the whole multisaddle region, supporting the assertion that multi-cut solutions really do dominate in this region and that the instanton condensation curves demarcate genuine phase transitions.}
    \label{fig:g-complex-regions}
\end{figure}

\begin{figure}
    \centering
    \begin{subfigure}[t]{0.4\textwidth}
        \centering
        \includegraphics[width=\linewidth, keepaspectratio]{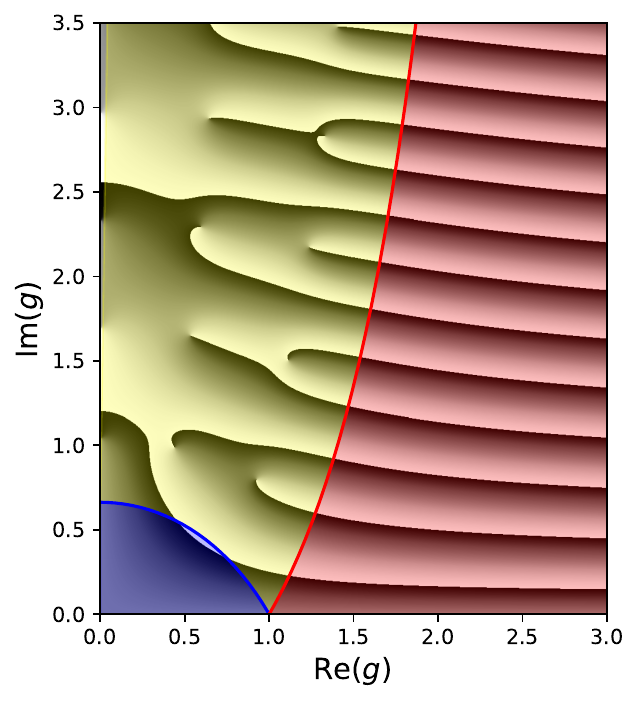}
        \caption{}
        \label{fig:toeplitz-complex-N5}
    \end{subfigure}
    \begin{subfigure}[t]{0.4\textwidth}
        \centering
        \includegraphics[width=\linewidth, keepaspectratio]{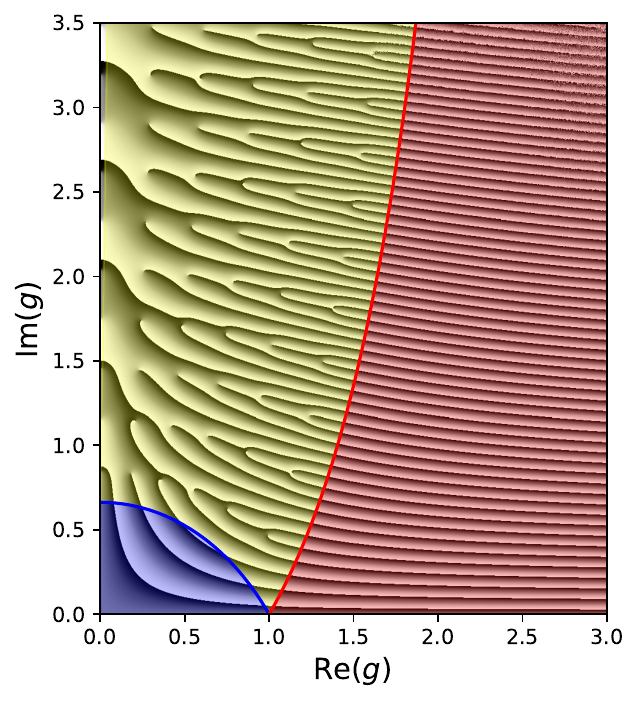}
        \caption{}
        \label{fig:toeplitz-complex-N11}
    \end{subfigure}
    \caption{Numerical evaluation of the Toeplitz determinant \eqref{eq:Toeplitz}, the finite-N analogue of the GWW partition functions, for (\subref{fig:toeplitz-complex-N5}) $N=5$ and (\subref{fig:toeplitz-complex-N11}) $N=11$. The colour corresponds to the complex argument, with the lightest and darkest colours of each region corresponding to $-\pi$ and $\pi$, respectively. As in Fig.~\ref{fig:g-complex-regions}, the blue, yellow and red regions correspond to dominance at large-$N$ by single-cut ungapped, multi-cut, and single-cut gapped solutions, respectively. The Lee-Yang zeros can be identified at the end of the branch points, where the colour cycles around all colours a point, showing the complex argument cycling anticlockwise by $2\pi$ around that point. We observe the Lee-Yang zeros filling the entire multi-cut region, with the edge of the Lee-Yang zeros aligning with the large-$N$ instanton condensation curve between the gapped and multi-cut regions. We expect the zeros to densely fill the whole multi-cut region as we increase $N$. This implies that multi-cut configurations indeed dominate the description in the large $N$ limit after we pass the instanton condensation point.}
    \label{fig:toeplitz-complex}
\end{figure}

Unlike the case of real coupling, we can no longer conclude that single-cut distributions will always be dominant. Multi-cut saddles might begin to dominate the free energy in regions where the sign of the single-cut instanton actions \eqref{eq:instanton-G} or \eqref{eq:instanton-UG} turn negative. The phase diagram for the $p=1$ model at complex coupling is shown in Fig.~\ref{fig:g-complex-regions}, and also appears in \cite{Copetti:2020dil}. Clearly, there are large regions in the complex plane where the instanton action becomes negative, where multi-cut solutions are expected to dominate.

A useful expression exists for the finite-N analogue of the $p=1$ truncation, or Gross-Witten-Wadia model. This is the Toeplitz determinant,
\begin{equation} \label{eq:Toeplitz}
    Z(g,N)=\mathrm{det}~\left[ I_{j-k} \left(gN\right) \right],
\end{equation}
where $I_{j-k}$ is the modified Bessel function and the indices $j$ and $k$ label the rows and columns of the matrix \cite{Rossi:1996hs,Bars:1979xb, Goldschmidt:1979hq}. This allows us to compute the partition function numerically at finite $N$. Evaluating this at complex values of the coupling $g$ at large finite $N$, we can see that there is indeed an apparent phase change in the region around the large-N instanton condensation boundaries. In particular, we observe that Lee-Yang zeros permeate the space where we expect multi-cut saddles to dominate \cite{Kolbig:1981qz}. As we increase $N$, they appear to fill the entire region in the complex plane between the two large-N instanton condensation curves. More precisely, in each the upper and lower half plane, there are $\frac{N}{2}$ trajectories of zeros. These seem to correspond to the $\frac{N}{2}$ pairs of eigenvalues that condense as we move across this region of the complex plane. Lee-Yang zeros generally indicate phase transitions, and are known to occur along anti-Stokes lines \cite{Itzykson:1983gb, pisani1993lee}. Each case of eigenvalue tunnelling is a type of anti-Stokes transition if looked at in $1/N$ resolution. We evaluate \eqref{eq:Toeplitz} over the complex plane for $N=5$ and $N=11$ and display it in Fig.~\ref{fig:toeplitz-complex}. The zeros are evidently aligning with the large-N instanton condensation lines.

In summary, we can be confident that multi-cut configurations dominate when the instanton action turns negative and that this corresponds to a genuine phase transition. There are also hints at interesting $1/N$ effects in these regions associated with tunnelling of individual eigenvalues. The instability we find for the BPS black hole is essentially captured by this $p=1$ model and the transition from single cut to multi-cut distributions.

\subsection{Higher truncations and classifying pinching points} \label{sec:higher-truncations}
Higher truncations exhibit increased complexity. The equations described in Sec.~\ref{sec:singel-cut-solution-definition} become more complicated, and more solutions become admissible. Multiple such solutions are expected to work together to give the full description of the black hole. Moreover, additional pinching points are introduced. As is obvious from \eqref{eq:y-general}, each increase in the truncation level introduces an additional pair of pinching points, labelled by $d_i$ and $d_i^{-1}$ inside the product. These differ in nature from the pinching point at $z=-1$ as their positions in the complex plane move around as the fugacity, $x$, is varied. The values of the endpoints $a$ and the additional pinching points $d_i$ can be calculated at any truncation  for any value of the complex fugacity, $x$, as described in Appendix~\ref{appendix:higher-truncations}. The explicit expressions for $p=2$ and $p=3$ are also provided in the appendix.

These new pinching points play two important roles. First, for each new pair of $d$-type pinching points, there is a new instanton corresponding to eigenvalues tunnelling as a pair to these points. We shall refer to them as $d$-type instantons. The instantons that tunnel to $z=-1$ will be referred to as $\pi$-type instantons.

Second, if the pinching points move across the single cut, the cut can split. Then, the assumption of a single cut, vital for the validity of the construction in Sec.~\ref{sec:singel-cut-solution-definition}, is violated.
This is very non-trivial to check and must be done numerically in general.

It is not obvious whether configurations that correspond to eigenvalue tunnelling to the new pinching points actually contribute to the original partition function. The most important instanton for us will continue to be the $\pi$-type instanton. We can be confident that the $\pi$-type instanton continues to contribute to the partition function in the same way in the higher truncations since higher truncations in our domain of interest can be described as perturbations on top of the $p=1$ model, for which we have strong and direct numerical confirmation of dominant multi-cut saddles. However, a full Lefschetz thimble analysis would be required to confirm that the new $d$-type instantons also contribute and lead to new multi-cut solutions that contribute to the original contour. 

The $d$-type pinching points can be classified into two types. This is easiest to define initially at real coupling. In the first case, the pinching point and its inversion sit on the unit circle, and we describe the pair as \emph{on-circle}. In the second, the pair lies more generally in the complex plane, one point lying inside the circle and its inverted pair outside. These will be described as \emph{off-circle}. At complex coupling, such a clear condition is harder to define (although it might be achieved by examining discriminants inside the radicals of expressions for the pinching points, such as those given in Appendix \ref{appendix:higher-truncations}). Instead, we classify pinching points as on-circle or off-circle if they are continuously connected to on-circle or off-circle configurations as we go to real coupling without any collision of the points as this trajectory is taken. In the very small black hole limit (i.e. the limit of small charge, $q \rightarrow 0$, at $O(N^2)$ level), all couplings become real. The on-circle complex pinching points then approach the unit circle. These points therefore lie close to the original contour over the circle as we approach the very small black hole. Eigenvalue tunnelling to $d$-type pinching points can also exhibit the phenomenon of ghost instantons, especially for off-circle pairs. In practice, we must confirm the sign of the instanton action independently.

\section{Numerical results for BPS black hole} \label{sec:numerical-results}
We will now present our results for the numerical analysis of the symmetric single-cut black hole saddle points and instanton actions. We will examine the first three truncations of the matrix model. The self-consistent single-cut saddles for these truncated models were identified in \cite{choi2024yang}, according to the procedure reviewed in Sec.~\ref{sec:index}. We will argue that there is a thermodynamic instability in these models against multi-cut solutions corresponding to instanton condensation along the small black hole branch. We will also find some tentative evidence for additional structure revealed only at higher truncations, where a different kind of instability might arise.

\subsection{$p=1$}
The first truncation considers only the first moment in the eigenvalue density distribution. It can be solved semi-analytically, as first demonstrated in \cite{Copetti:2020dil}, where the physically relevant saddle was found to be given by,
\begin{equation}
    g_{\star} = a_1 + \sqrt{a_1(a_1-1)}.
\end{equation}
This then has the effective free energy,
\begin{equation}
    Q = \frac{1}{2} \left(a_1 + \sqrt{a_1(a_1-1)} - \log \left(a_1 + \sqrt{(a_1(a_1-1))} \right) - 1 \right).
\end{equation}
The contours $\mathrm{Re}~Q=0$ determine the deconfinement curve, or Hawking-Page transition, in the complex fugacity plane.

As described in Sec.~\ref{sec:BPS-black-holes}, to obtain the physical black hole saddle, we perform the Legendre transform to find the entropy,
\begin{equation}
    S = Q - q \log x
\end{equation}
where we have charge determined by,
\begin{equation}
    q = \frac{\partial Q}{\partial x} x
\end{equation}
with the condition on the physical saddle that $q$ is real. 
This saddle was found previously in \cite{choi2024yang}.

We then check the thermodynamic stability of this solution by computing the one-instanton action \eqref{eq:instanton-G} at every point along the saddle. This instanton action corresponds to a single eigenvalue tunnelling to the pinching point at $z=-1$. This is a $\pi$-type instanton, no $d$-type instantons existing at $p=1$.  A negative instanton action would suggest that the opening of a new cut at $z=-1$ is thermodynamically favoured, and that the superconformal index should be dominated instead by a multi-cut distribution.

\begin{figure}[htbp]
    \centering
    \includegraphics[width=0.8\textwidth]{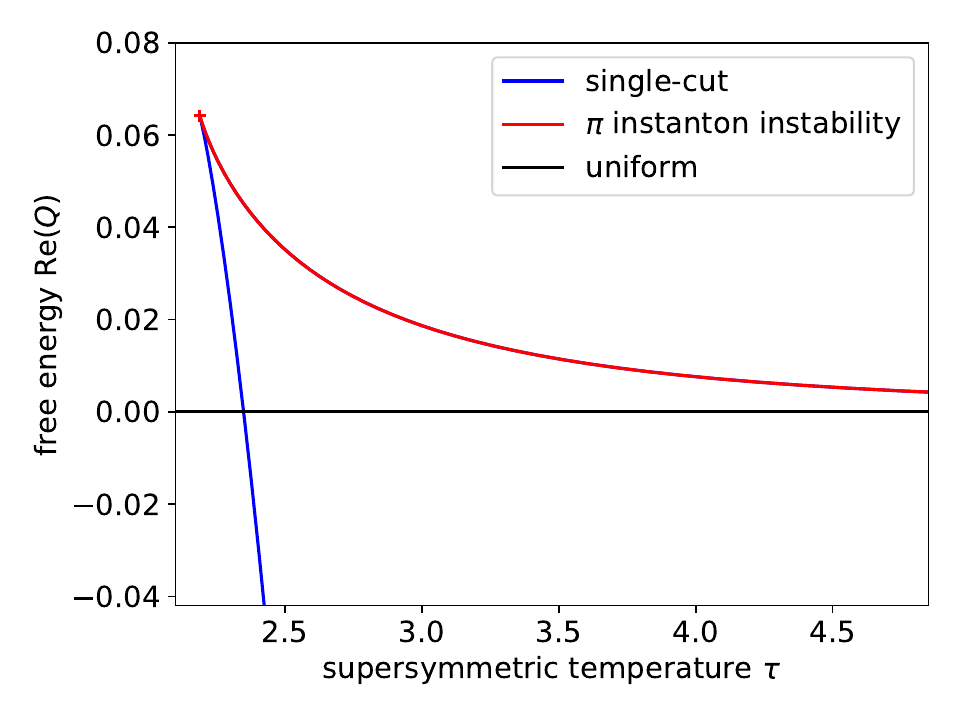} 
    \caption{Free energy and location of the instanton instability for the black hole saddle in the $p=1$ truncation. Investigations at high precision show that the instability does not lie along the entire negative heat capacity saddle, despite appearances here.}
    \label{fig:p1-free-energy}
\end{figure}

\begin{figure}[htbp]
    \centering
    \begin{subfigure}[b]{0.45\textwidth}
        \includegraphics[width=\textwidth]{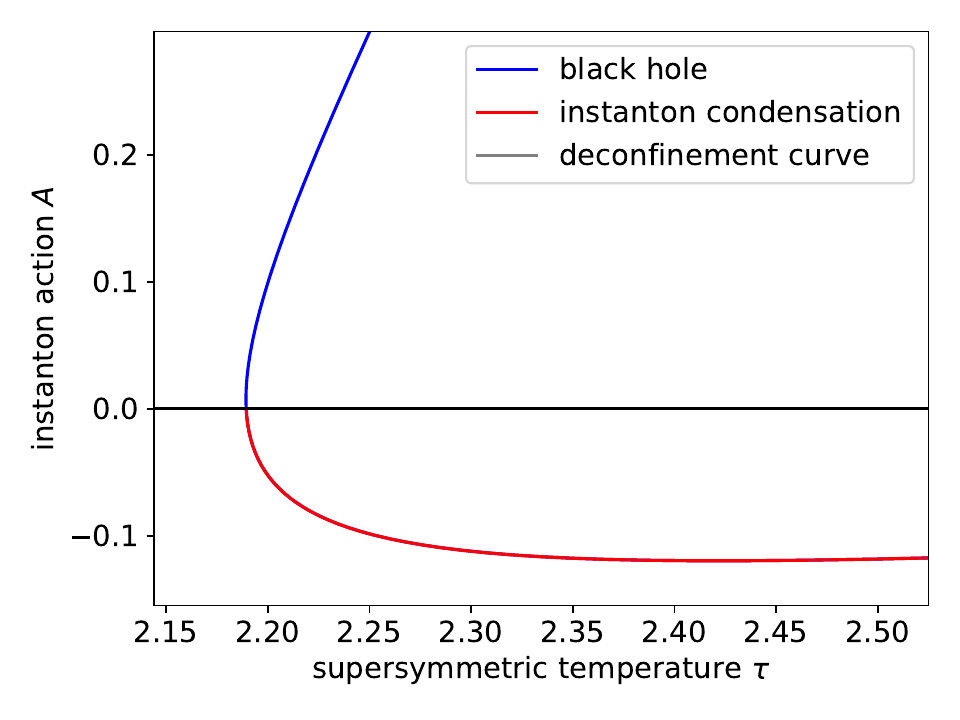}
        \caption{}
        \label{fig:p1-MCE-instanton-action}
    \end{subfigure}
    \hfill
    \begin{subfigure}[b]{0.45\textwidth}
        \includegraphics[width=\textwidth]{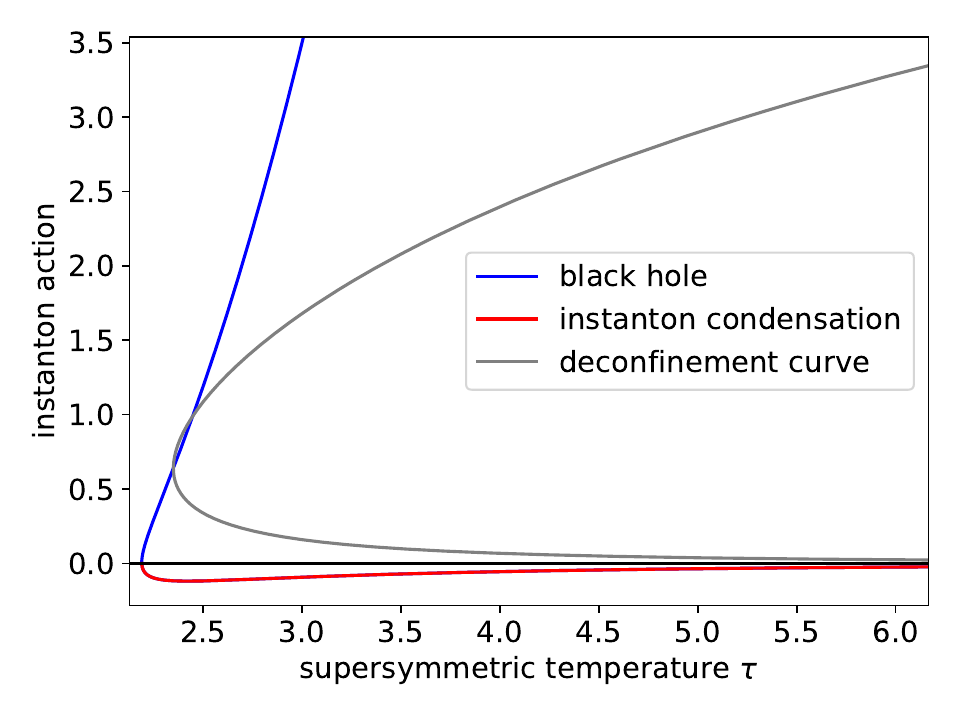}
        \caption{}
        \label{fig:p1-GCE-instanton-action}
    \end{subfigure}
    \caption{(\subref{fig:p1-MCE-instanton-action}) Instanton tunnelling action for the black hole saddle in the $p=1$ model in the microcanonical ensemble, against supersymmetric temperature. The lower, condensed region lies along the small black hole branch. (\subref{fig:p1-GCE-instanton-action}) Zoomed out and now including the deconfinement curve at $p=1$, which shows no instanton condensation.}
    \label{fig:p1-instanton-action}
\end{figure}

\begin{figure}[htbp]
    \centering
    \begin{subfigure}[b]{0.3\textwidth}
        \includegraphics[width=\textwidth]{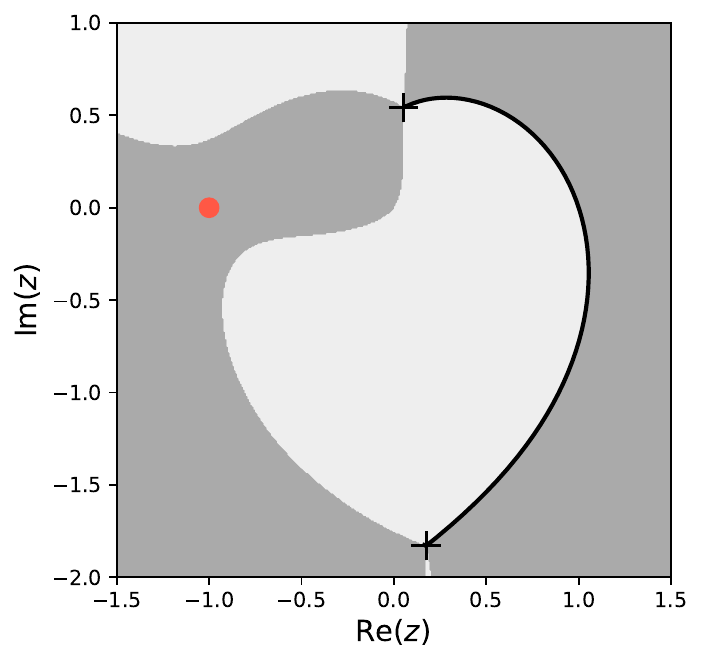}
        \caption{}
        \label{fig:yint-p1-after-cond}
    \end{subfigure}
    \hfill
    \begin{subfigure}[b]{0.3\textwidth}
        \includegraphics[width=\textwidth]{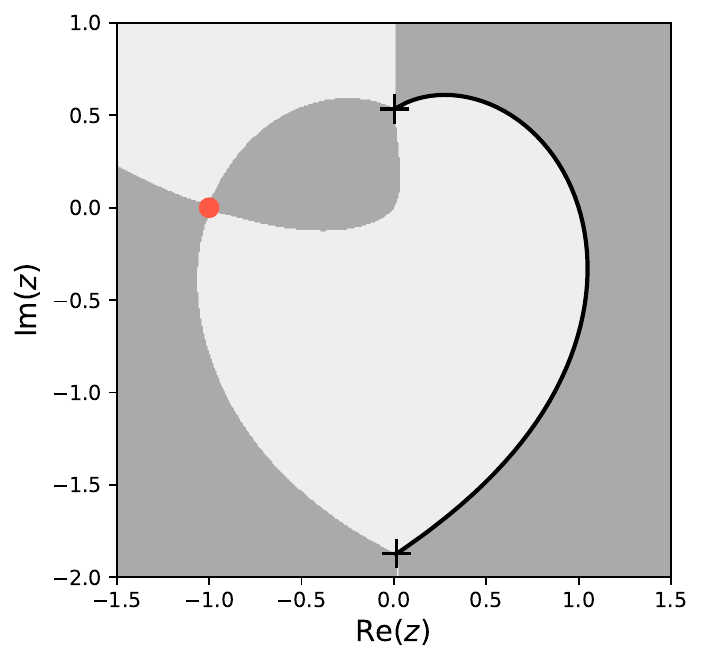}
        \caption{}
        \label{fig:yint-p1-at-cond}
    \end{subfigure}
    \hfill
    \begin{subfigure}[b]{0.3\textwidth}
        \includegraphics[width=\textwidth]{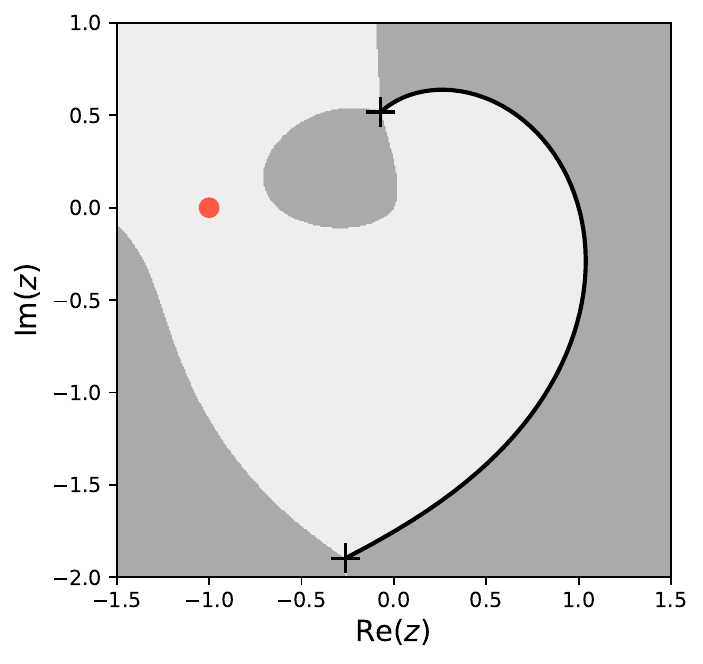}
        \caption{}
        \label{fig:yint-p1-before-cond}
    \end{subfigure}
    \hfill
    \caption{The eigenvalue contour and electric potential in the $p=1$ truncation around the instanton condensation point. Dark grey indicate positive action for eigenvalues tunnelling from the occupied cut (positive effective electric potential \eqref{eq:Veff-integral}), while light grey indicates negative action (negative effective potential). The red dot indicates the pinching point at $z=-1$. This shows the points (\subref{fig:yint-p1-after-cond}) before condensation at $q=1.30$ ($\tau=2.20$), (\subref{fig:yint-p1-at-cond}) at the point of condensation at $q=1.04$ ($\tau=2.19$), and (\subref{fig:yint-p1-before-cond}) after the condensation at $q=0.73$ ($\tau=2.22$). The $z=-1$ pinching point enters the area of negative potential, showing that the configuration becomes unstable against $\pi$-type instanton condensation.}
    \label{fig:yint-p1}
\end{figure}

The results are shown in Figs.~\ref{fig:p1-free-energy} and \ref{fig:p1-instanton-action}.
We see that condensation of the $\pi$-instanton appears along almost the entire small black hole saddle. Careful numerics show that the condensation actually takes place slightly below the cusp. 
Along the deconfinement curve, we see no such instanton condensation. Instead, the instanton action smoothly decreases to zero from above as the very small black hole limit is approached. The instability is thus shown to be a property of the microcanonical saddle and the physical BPS black hole. 
The eigenvalue distribution and effective potential are shown around the instanton condensation point in Fig.~\ref{fig:yint-p1}.

\begin{figure}[htbp]
    \centering
    \includegraphics[width=0.7\textwidth]{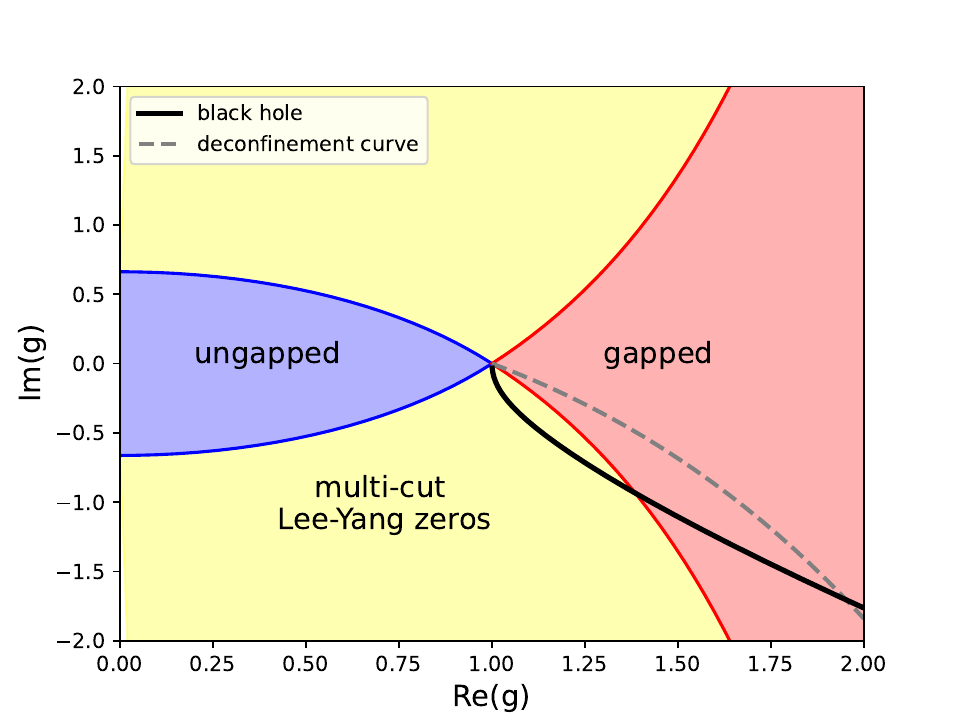} 
    \caption{The path of the GWW coupling $g$ taken by the physical black hole saddle (solid, black) and the deconfinement curve (dashed, grey) in the $p=1$ model. The blue and red lines correspond to instanton condensation lines, separating the single-cut regions from the multi-cut region in yellow, as in Fig.~\ref{fig:g-complex-regions}. The black hole saddle crosses the instanton condensation point region, indicating an instability of the single-cut solutions towards multi-cut distributions. The deconfinement curve remains within the single-cut region.}
    \label{fig:g-path}
\end{figure}

It is natural at this point to wonder whether the instability that appears according to this calculation merely represents the well-known and obvious instability in the grand canonical ensemble that accompanies any saddle with negative heat capacity. For example, one might suppose that the condensation of eigenvalues around $z=-1$ could be an alternative pathway to the uniform distribution, which could be the endpoint of this instability. 
To become convinced that this instability is instead something distinct, we should recall what the Legendre transform achieves on saddles. Taking the Legendre transform provides us with a thermodynamic potential (the entropy), $S(q,\nu)$, whose derivative with respect to $x$ now vanishes. The charge $q$ is tuned to cancel derivatives with respect to $x$ at $\nu=0$, and the changes from a small tunnelling fraction $\delta \nu$ lead to differences only at quadratic order. Therefore, the instability found by evaluating the sign of the instanton action at fixed $q$ is a genuine instability of the in the microcanonical ensemble. To be explicit, we compare the entropy at equal charge of two saddles differing by eigenvalue tunnelling fraction $\delta \nu$, which is held fixed in each.  We will introduce the free energy $Q(x,\nu)$ for saddles where a fraction $\nu$ (continuously valued) of the eigenvalues are tunnelling. The instanton action can be defined as $A(x)=\partial_{\nu}Q(x,\nu)\lvert_{\nu=0}$. We take $x+\delta x$ as the fugacity of the second saddle at the given charge $q$. Then we find, 
\begin{eqnarray}
    S(q,0) &=& Q(x,0) - q \log x \nonumber \\
    S(q,\delta\nu) &=& Q(x + \delta x,\delta\nu) - q \log (x + \delta x) \nonumber  \\
    & \approx & Q(x,0) + \frac{\partial Q(x,0)}{\partial x} \delta x + \delta \nu A - q \log x - q \frac{\delta x}{x} \nonumber  \\
    \implies & & S(x,\delta\nu) - S(x,0) \approx \frac{\partial Q_0(x)}{\partial x} \delta x + \delta \nu A - q \frac{\delta x}{x} \nonumber ,
\end{eqnarray}
and using the definition $q=\frac{\partial Q(x,0)}{\partial x} x$, we see that the favoured saddle in the microcanonical ensemble depends only the sign of the instanton action, $A$. Therefore, checking the sign of the instanton action is enough to determine an instability against eigenvalue tunnelling in the microcanonical ensemble.

Of course, it is possible that the instability is an artefact of the truncation, vanishing in the exact calculation. The approximation becomes less accurate as we move to smaller charges, i.e. as the size of the black hole shrinks. 
This provides motivation to look at higher order corrections. 
However, we mention briefly now that the numerical values of $a_n/n$, which control the strength of the correction terms, support the first truncation being a good approximation of the full theory in the area around the onset of the instability. We will wait until we examine the $p=3$ truncation to develop this line of argument further.

\subsection{$p=2$}

\begin{figure}[htbp]
    \centering
    \begin{subfigure}[b]{0.45\textwidth}
        \includegraphics[width=\textwidth]{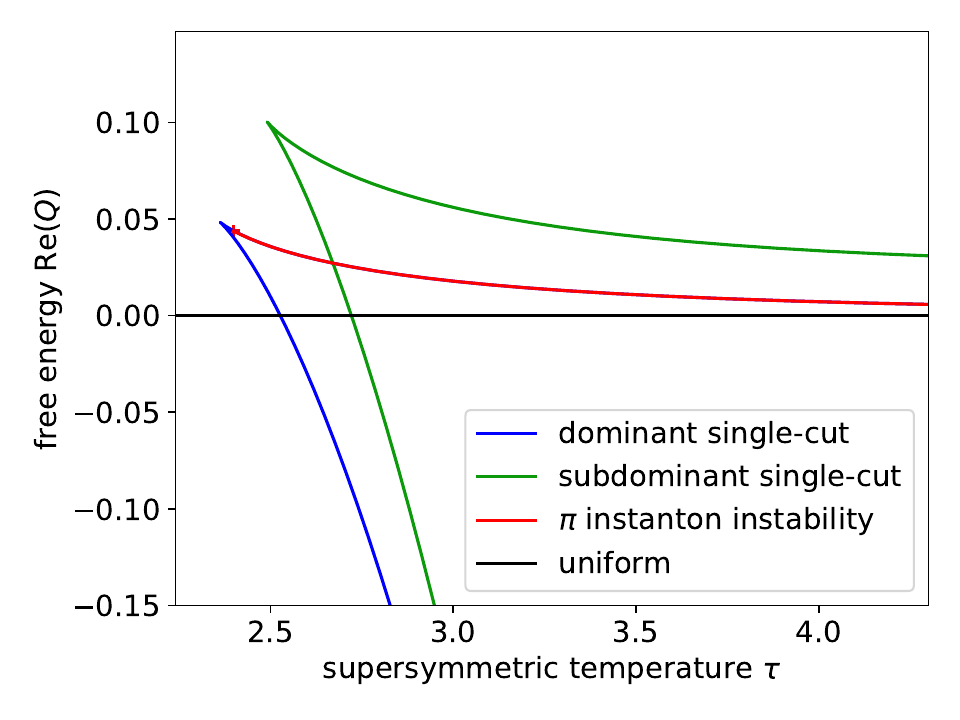}
        \caption{}
        \label{fig:p2-free-energy}
    \end{subfigure}
    \hfill
    \begin{subfigure}[b]{0.45\textwidth}
        \includegraphics[width=\textwidth]{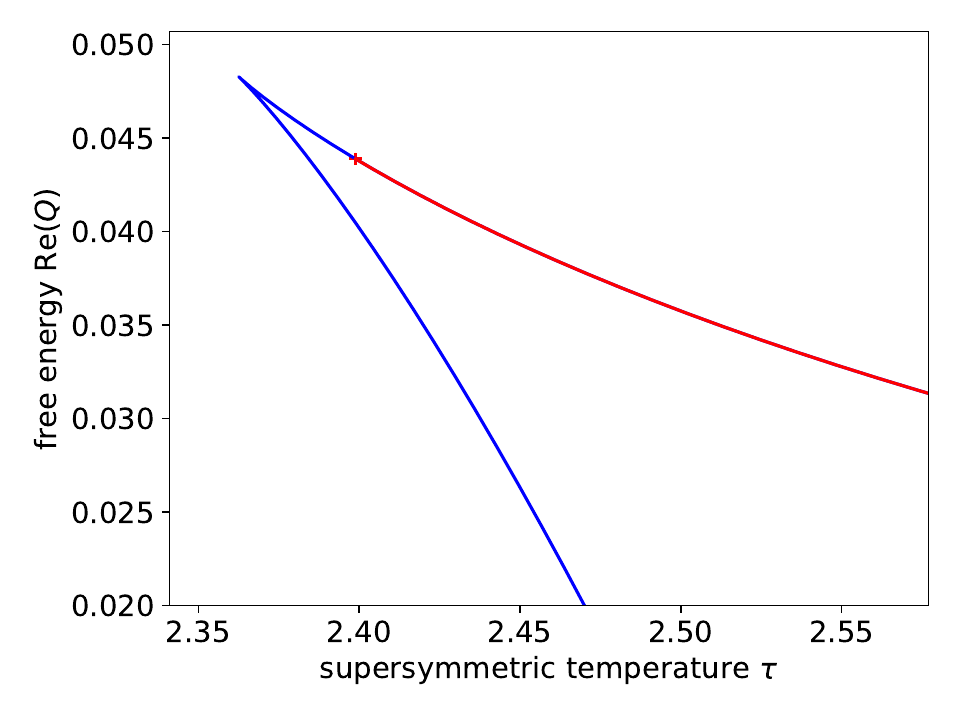}
        \caption{}
        \label{fig:p2-free-energy-zoomed}
    \end{subfigure}
    \caption{(\subref{fig:p2-free-energy}) The free energies of the two saddles that together describe the BPS black hole in the $p=2$ model. By examining the entropy at fixed charge, we find that the saddle depicted by the blue curve is dominant in the microcanonical saddle in the region around the instability. This saddle is found to be unstable against $\pi$-type instantons along the small black hole region, just below the cusp. The subdominant saddle exhibits no instanton instabilities. (\subref{fig:p2-free-energy-zoomed}) Zoomed in around the region of the cusp}
    \label{fig:p2-results}
\end{figure}

When we truncate the model to two modes, $p=2$, the situation becomes more complex. There are now two consistent single-cut saddles that can represent the standard black hole. Both of these saddles are thought to be important in the full description of the black hole \cite{choi2024yang}. The results are shown in Fig.~\ref{fig:p2-results}. We see that one of these saddles exhibits an instability against instanton condensation to the $\pi$-type instanton. This is very similar to the instability found at $p=1$, although it has withdrawn down the small black hole saddle somewhat, taking place at lower charge than the cusp. By examining the entropies of the two single-cut saddles as a function of charge, we find that the saddle exhibiting the $\pi$-tunnelling instability is thermodynamically dominant over the other single-cut saddle in the microcanonical ensemble. The true dominant configuration in the microcanonical ensemble at this level of truncation should therefore be associated with the endpoint of the instability.

The subdominant saddle does not have an analogous instability. 
Along the small black hole, we find that neither saddle exhibits an instability against eigenvalue tunnelling to the $d_1$ pinching point, which is always `off-circle' along both saddles, according to the terminology of Sec.~\ref{sec:higher-truncations}. 

\subsection{$p=3$} \label{sec:results-p3}

\begin{figure}[htbp]
    \centering
    \includegraphics[width=0.8\textwidth]{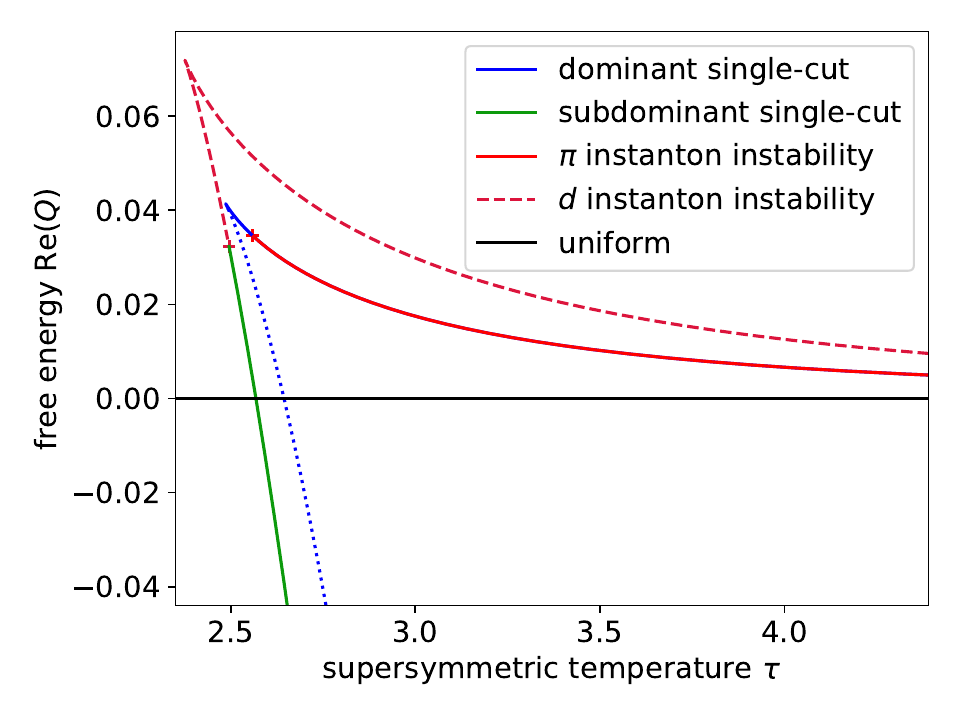} 
    \caption{Free energy of the two self-consistent single-cut saddles at $p=3$ truncation. The blue saddle, which is dominant in the microcanonical ensemble as long as it exists, shows an instability against $\pi$-type instantons. The dotted blue line indicates where this saddle fails to be self-consistent. The `subdominant' saddle, shown in green, may also show an instability against eigenvalue tunnelling to one of the $d$-type pinching points along the dashed crimson region.}
    \label{fig:p3-free-energy}
\end{figure}

\begin{figure}[htbp]
    \centering
    \begin{subfigure}[b]{0.45\textwidth}
        \includegraphics[width=\textwidth]{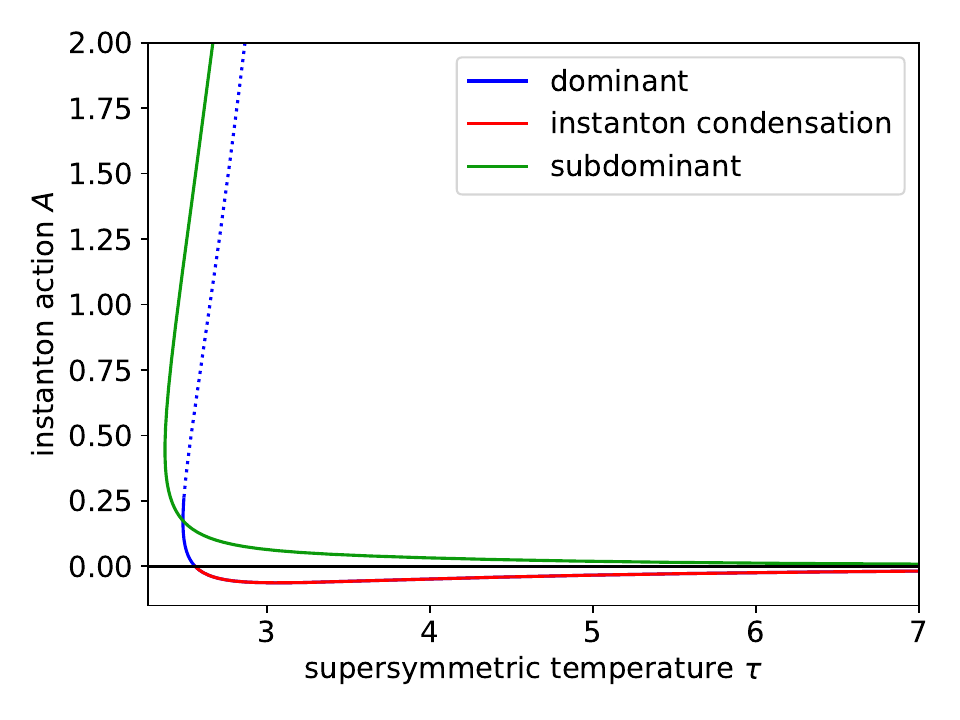}
        \caption{}
        \label{fig:p3-instanton-action}
    \end{subfigure}
    \hfill
    \begin{subfigure}[b]{0.45\textwidth}
        \includegraphics[width=\textwidth]{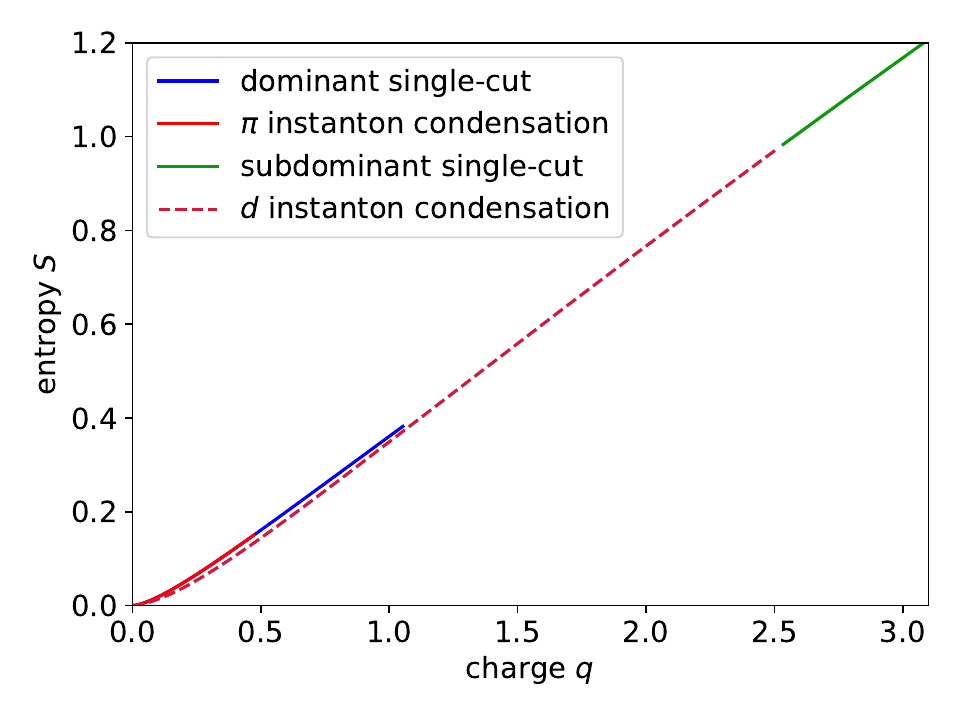}
        \caption{}
        \label{fig:p3-entropy}
    \end{subfigure}
    \caption{(\subref{fig:p3-instanton-action}) Instanton action for $\pi$-type instantons for the two single-cut saddles in the physically-relevant region of the $p=3$ truncation against supersymmetric temperature. The dashed blue region shows where the dominant saddle fails to admit a consistent single-cut eigenvalue distribution and is therefore invalid. (\subref{fig:p3-entropy}) The entropy of each saddle against charge. The `dominant' saddle, described by the blue curve, has greater entropy at all charges for which it exists, and in particular at the onset of the $\pi$-type instability.
    The `subdominant' saddle is shown in green. Where the `dominant' saddle ceases to exist, this `subdominant' saddle is the only viable single-cut saddle, and therefore becomes dominant saddle of the microcanonical description. 
    This saddle exhibits a potential instability against condensation of $d$-type instantons, including in a region of charge $q$ for which it is the only self-consistent single-cut saddle solution.
    }
    \label{fig:p3-instanton-action-and-entropy}
\end{figure}

For the $p=3$ truncation, again, two single-cut saddles are physically relevant to the description of the black hole. As noted in \cite{choi2024yang}, some phenomena expected to be exhibited by the full BPS description in the small black hole region appear only at $p \geq 3$, giving an additional reason to examine this truncation level. 

\begin{figure}[htbp]
    \centering
    \begin{subfigure}[b]{0.3\textwidth}
        \includegraphics[width=\textwidth]{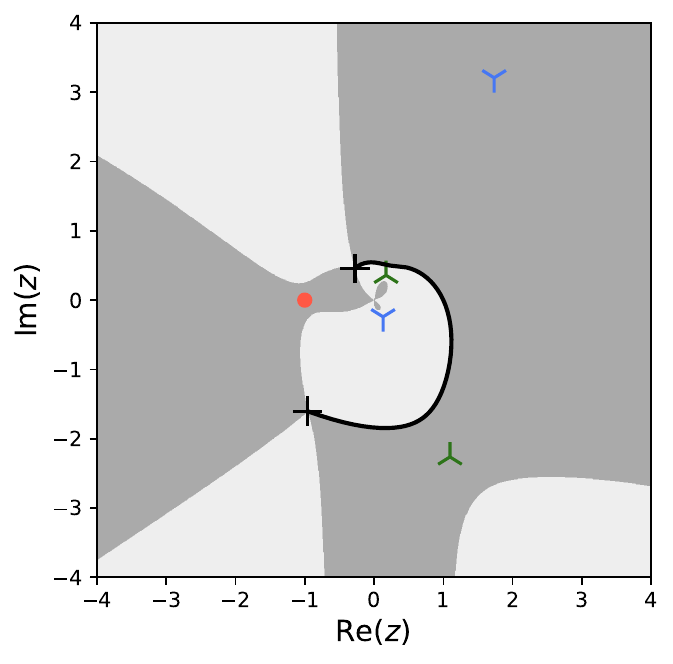}
        \caption{}
        \label{fig:yint-saddle2-i800}
    \end{subfigure}
    \hfill
    \begin{subfigure}[b]{0.3\textwidth}
        \includegraphics[width=\textwidth]{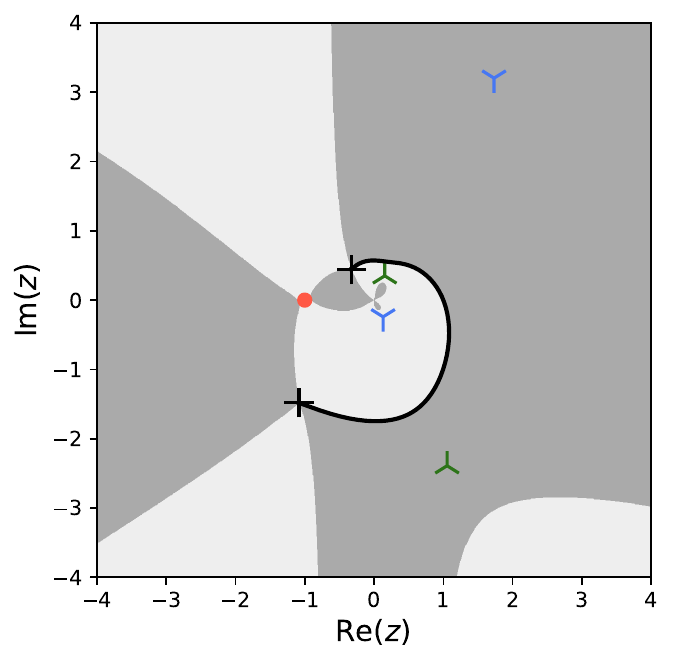}
        \caption{}
        \label{fig:yint-saddle2-i740}
    \end{subfigure}
    \hfill
    \begin{subfigure}[b]{0.3\textwidth}
        \includegraphics[width=\textwidth]{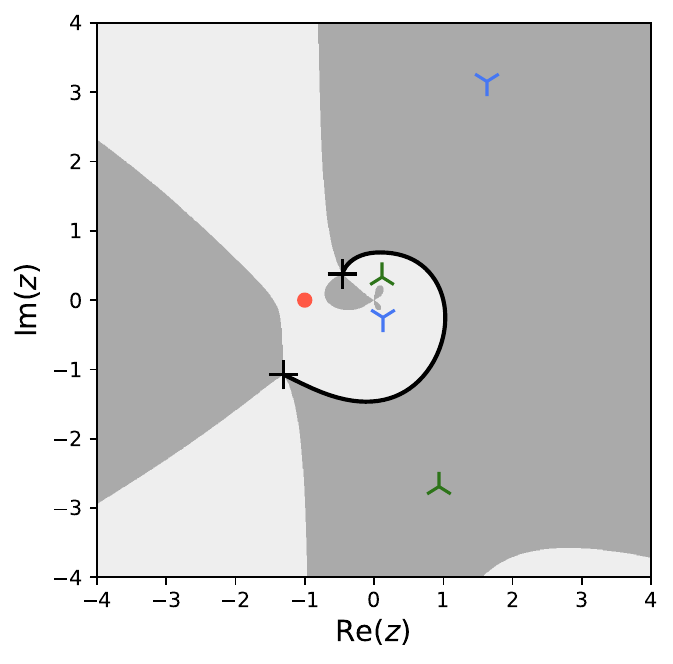}
        \caption{}
        \label{fig:yint-saddle2-i710}
    \end{subfigure}
    \caption{The eigenvalue contour and pinching point locations in the $p=3$ truncation along the dominant saddle around the $\pi$-instanton condensation point. The black curve indicates the cut occupied by eigenvalues, with the black crosses giving the endpoints. The red dot is the pinching point at $z=-1$, while the blue and green triangular ticks markers give the locations of the $d$-type pinching points and their inverses. The sign of the effective electric potential at all points is given by the background colour, with dark and light grey denoting positive and negative potential, respectively. The configurations show the points (\subref{fig:yint-saddle2-i800}) before condensation at $q=0.64$ ($\tau=2.51$), (\subref{fig:yint-saddle2-i740}) at the point of condensation at $q=0.46$ ($\tau=2.57$), and (\subref{fig:yint-saddle2-i710}) after the condensation at $q=0.18$ ($\tau=3.10$). The $z=-1$ pinching point enters the area of negative potential, showing that the configuration becomes unstable against $\pi$-type instanton condensation. Note that the pinching points found in the light grey region partially encircled by the cut correspond to ghost instantons and do not indicate an instability.}
    \label{fig:yint-saddle2}
\end{figure}

The saddle with dominant entropy in the small black hole region is again sensitive to $\pi$-instanton condensation against tunnelling to the pinching point at $z=-1$, as shown in Figs.~\ref{fig:p3-free-energy} and \ref{fig:p3-instanton-action-and-entropy}. It has again shifted slightly further down the small black hole branch, but still lies close to the cusp. The configurations around the onset of the instability are shown in Fig.~\ref{fig:yint-saddle2}.
The single-cut solution for this saddle is destroyed when one of the pinching points passes across the cut, as already stated in \cite{choi2024yang}, and indicated by the termination of the solid blue line in Figs.~\ref{fig:p3-free-energy} and \ref{fig:p3-instanton-action-and-entropy}. Although perhaps not clearly visible in Fig.~\ref{fig:p3-free-energy}, we confirm that the cusp of the free energy of the dominant saddle lies fully within the self-consistent solution.

\begin{figure}[htbp]
    \centering
    \begin{subfigure}[b]{0.3\textwidth}
        \includegraphics[width=\textwidth]{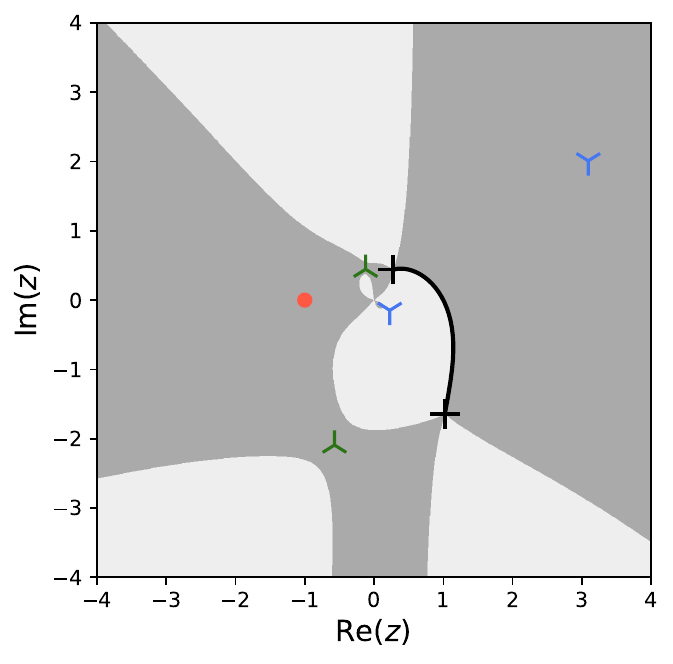}
        \caption{}
        \label{fig:yint-saddle1-i620}
    \end{subfigure}
    \hfill
    \begin{subfigure}[b]{0.3\textwidth}
        \includegraphics[width=\textwidth]{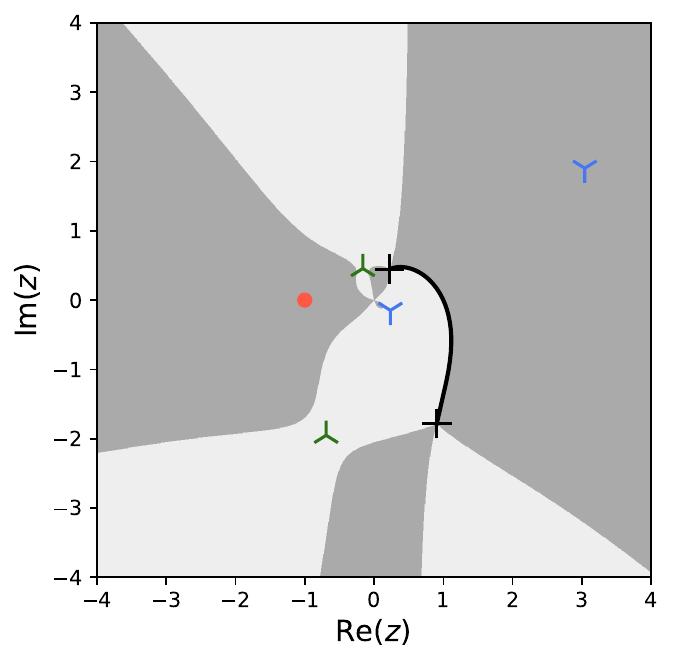}
        \caption{}
        \label{fig:yint-saddle1-i590}
    \end{subfigure}
    \hfill
    \begin{subfigure}[b]{0.3\textwidth}
        \includegraphics[width=\textwidth]{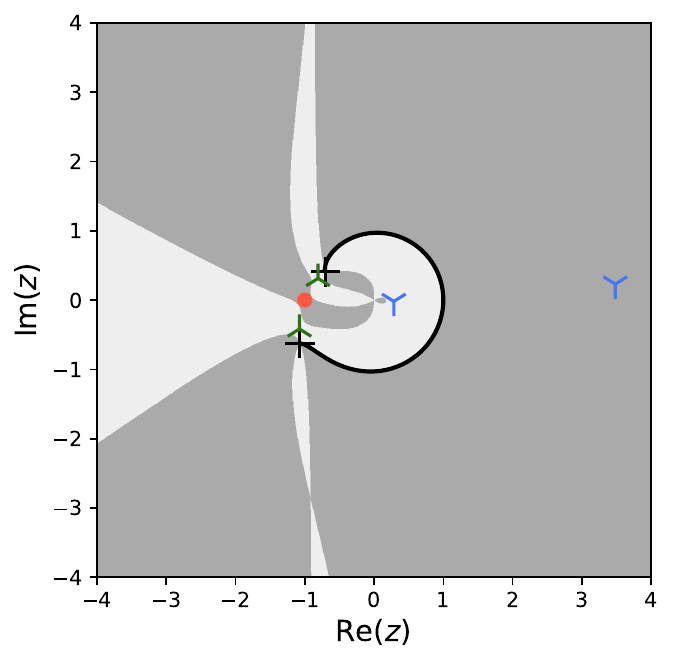}
        \caption{}
        \label{fig:yint-saddle1-i0}
    \end{subfigure}
    \caption{Locations of the pinching points in the complex plane at various points along the `subdominant' (lower entropy) saddle in the $p=3$ truncation, which is the only viable saddle in the microcanonical ensemble in this region and therefore actually dominant here. The configurations shown in (\subref{fig:yint-saddle1-i620}) at $q=3.00$ ($\tau=2.55$) and (\subref{fig:yint-saddle1-i590}) at $q=2.03$ ($\tau=2.44$) sit just before and after the apparent $d$-type instanton condensation. The pinching points indicated by the green triangular tick markers enter the light grey area corresponding to negative eigenvalue tunnelling action. This indicates a potential instability against eigenvalue tunnelling to these pinching points.
    (\subref{fig:yint-saddle1-i0}) shows the very small black hole limit at $q=10^{-3}$ ($\tau=47.5$), where the occupied contour is seen returning to the unit circle. It is clear from this that the green $d$-type pinching points are of \emph{on-circle} type. The pinching point denoted by the blue downward-pointing triangular tick marker that sits in the light grey region corresponds to a ghost instanton and does not indicate an instability.
    }
    \label{fig:yint-saddle1}
\end{figure}

We call this first saddle `dominant' because it has greater entropy in the region of greatest interest to us, the small black hole saddle and region around its $\pi$ instanton condensation. However, at charges above the point at which this saddle ceases to exist, what we call the `subdominant' saddle becomes the only self-consistent single-cut description of the black hole in the microcanonical ensemble. 

The second saddle has less entropy than the first saddle in the region of the latter's $\pi$ instanton condensation, and is therefore subdominant in the microcanonical ensemble in the region of most interest to us. However, there is a region of charge away from this $\pi$-instability for which the `dominant' saddle is not self-consistent. Interestingly, this second saddle exhibits an apparent instability against condensation of the new $d$-type instantons, including in regions where it is the only self-consistent single-cut saddle. This is depicted in Fig.~\ref{fig:p3-instanton-action-and-entropy}. This potential condensation first takes place while the black hole is still large, i.e. while it still has positive heat capacity, but still above the Hawking-Page transition in the grand canonical ensemble, as seen in Fig.~\ref{fig:p3-free-energy}. 

The configurations before and after this potential instability are shown in Fig.~\ref{fig:yint-saddle1}. Also evident in this figure, these instantons are on-circle in the vocabulary of Sec.~\ref{sec:higher-truncations}, corresponding to pinching points that move onto the unit circle in the very small black hole limit. This is the only saddle we have examined that has pinching points that are on-circle. It seems plausible that any on-circle pinching points exhibited at higher truncation levels would also enter the region of negative effective potential as we approach the small charge limit and appear to condense.

Whether it indicates a genuine instability along the saddle, i.e. whether it contributes to the original contour of integration in a way that would be revealed by a full Lefschetz thimble analysis, is left unresolved in this work, as is the question of any physical significance such an instability might carry. However, we note that an instability of this nature could only be discovered at sufficiently large truncations $p$ since it requires on-circle $d$-type pinching points to exist. In the discussion section, we propose a very speculative connection between this potential instability and the results from using a different method to calculate BPS black hole saddles, namely the Bethe ansatz approach \cite{Benini:2018ywd}. 

\begin{figure}[htbp]
    \centering
    \begin{subfigure}[b]{0.55\textwidth}
        \includegraphics[width=\textwidth]{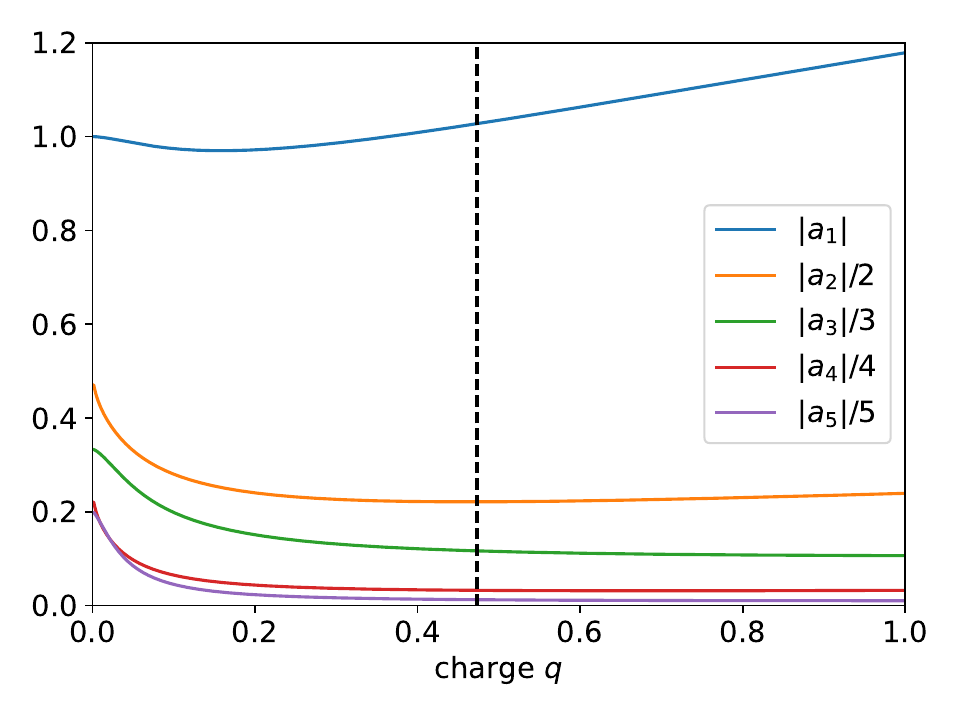}
        \caption{}
    \end{subfigure}
    \caption{Values of $|a_n|/n$ along the dominant $p=3$ saddles. The corrections to thermodynamic quantities at each truncation depend on $a_n/n$. 
    }
    \label{fig:p3-an}
\end{figure}

\begin{figure}[htbp]
    \centering
    \begin{subfigure}[b]{0.65\textwidth}
        \includegraphics[width=\textwidth]{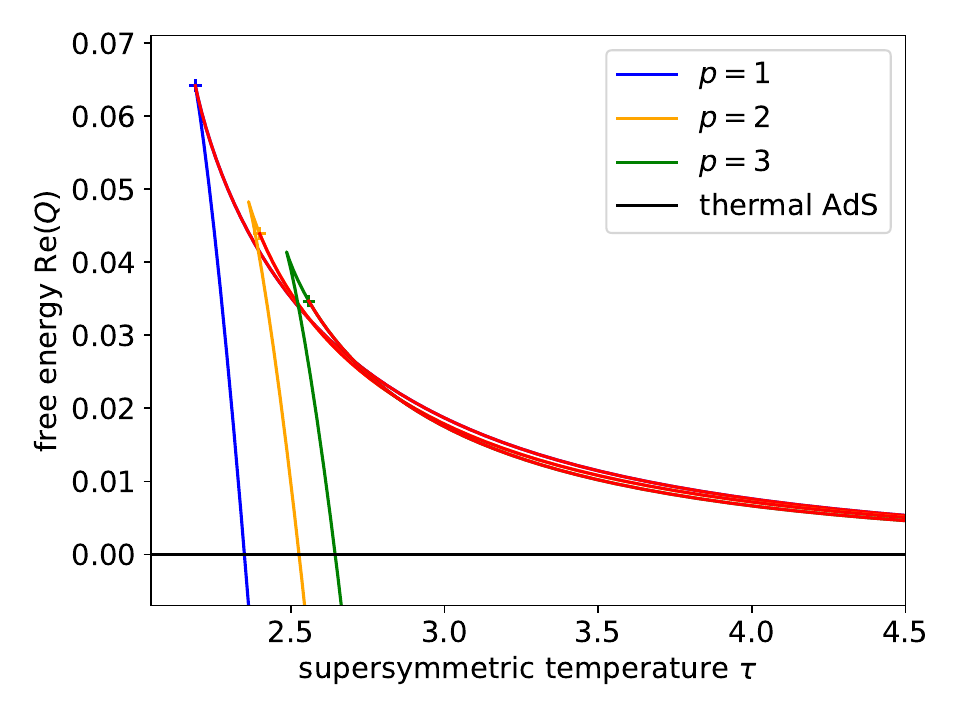}
    \end{subfigure}
    \caption{Free energy against supersymmetric temperature, $\tau$, for the dominant saddles in the microcanonical ensemble of the first three truncations, $p=1,2,3$, each showing the region of instability against $\pi$-type instanton condensation}
    \label{fig:all-dominant-free-energy}
\end{figure}

\begin{figure}[htbp]
    \centering
    \begin{subfigure}[b]{0.45\textwidth}
        \includegraphics[width=\textwidth]{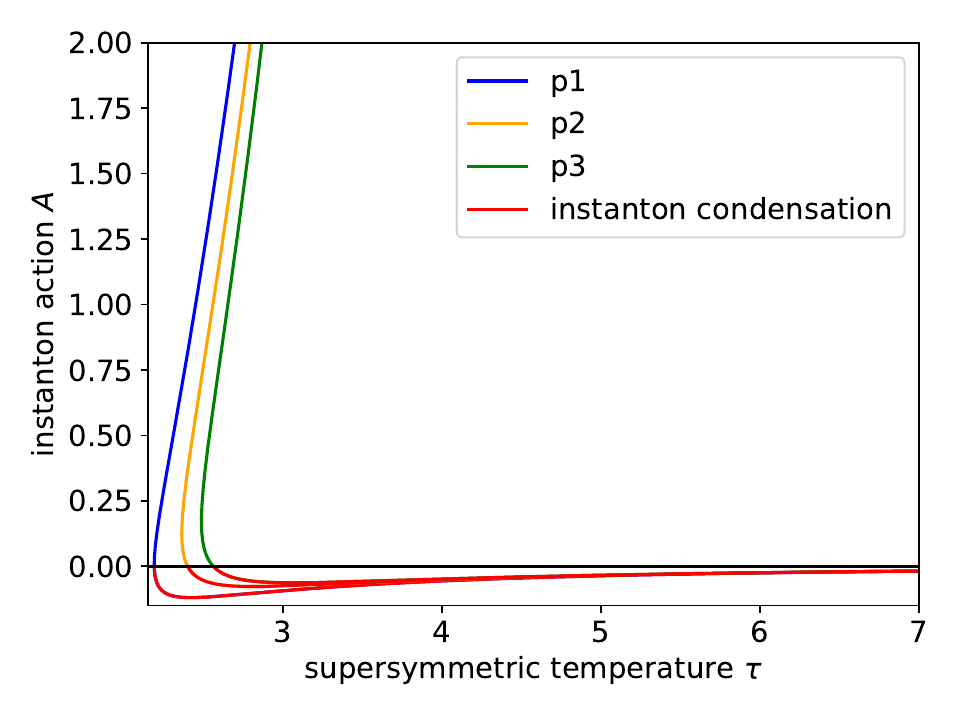}
        \label{fig:all-dominant-instanton-unzoomed}
    \end{subfigure}
    \hfill
    \begin{subfigure}[b]{0.45\textwidth}
        \includegraphics[width=\textwidth]{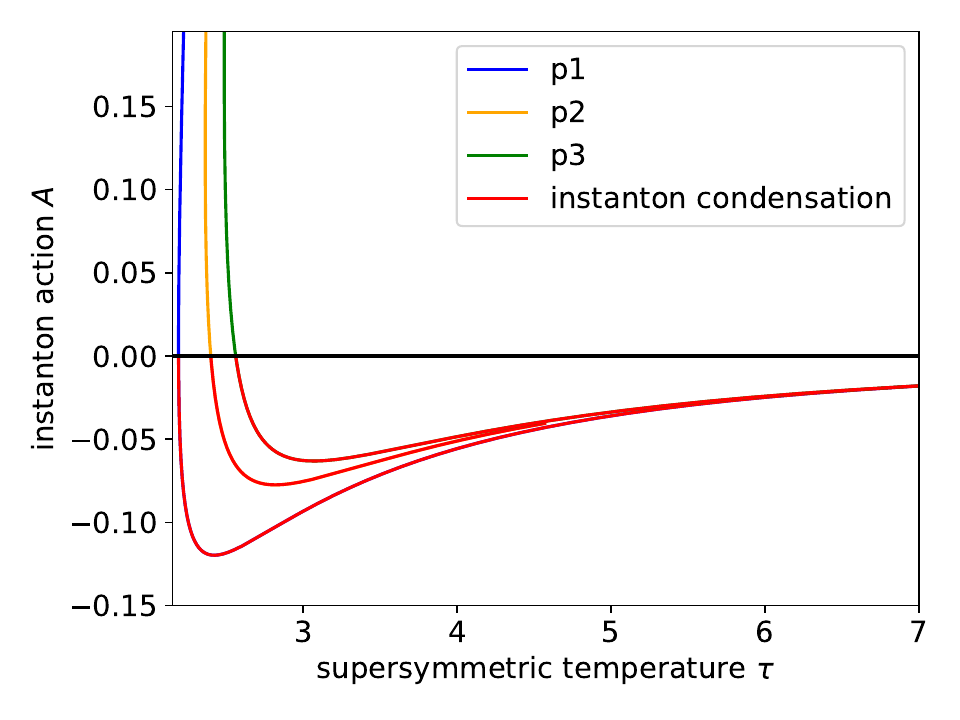}
        \label{fig:all-dominant-instanton-zoomed}
    \end{subfigure}
    \caption{
    Instanton actions for the $\pi$-type instantons against supersymmetric temperature, $\tau$, for the dominant saddles in the microcanonical ensemble of the first three truncations, $p=1,2,3$. 
    }
    \label{fig:all-dominant-instanton}
\end{figure}

Returning to our main topic of interest, we conclude that in some region close to and below the cusp, there are no self-consistent single-cut saddles that dominate the microcanonical ensemble. Multi-cut saddles are necessary for describing the black hole at this level of matrix truncation. The corrections to saddle entropies, integrals of $y(z)$, and other quantities relevant to the instanton tunnelling receive corrections proportional to $\frac{a_n}{n}$. The absolute values are shown for charges along the saddle in Fig.~\ref{fig:p3-an}. It is evident that corrections for $n>3$ are very small numerically in the region where we first find instanton condensation. In particular,
at $q \sim 0.47$, the point at which the $\pi$ instanton first condenses, we find $\frac{|a_4|}{4} \sim 0.032$. 
Therefore, it appears very likely that the $\pi$ instanton instability persists when further corrections to the matrix model are included, and that it reflects a true instability in the full BPS black hole saddle. A comparison of the instanton condensation regions for all entropy-dominating saddles of the first three truncation levels is given in Fig.~\ref{fig:all-dominant-free-energy}, while the corresponding $\pi$-instanton actions are shown in Fig.~\ref{fig:all-dominant-instanton}. 

\section{Discussion}\label{sec:discussion}

\subsection{Summary of results}

We have presented robust evidence that an instanton instability exists along the conventional Gutowski-Reall BPS black hole saddle in the microcanonical ensemble at sufficiently small charge. This implies that multi-cut distributions, corresponding to a new gravitational phase, dominate over the standard small black hole in this region. There is good evidence that this instability, driven by eigenvalues tunnelling to the pinching point at $z=-1$, appears in the exact BPS phase diagram, and is not a mere artefact of the truncations employed in our approximation. This is substantiated, for example, by the robustness of the instability under increasing truncations, as shown in Fig.~\ref{fig:all-dominant-free-energy} and Fig.~\ref{fig:all-dominant-instanton}, and of the magnitude of higher correction terms $|a_n|/n$, shown in Fig.~\ref{fig:p3-an}. We can also be confident that the corresponding multi-cut solutions resulting from $\pi$-type instanton condensation indeed contribute to the original contour of integration. The $p=1$ truncation can be evaluated directly using the Toeplitz determinant description \eqref{eq:Toeplitz}, which indicates a phase transition where we expect $\pi$-instanton condensation to take place. Higher truncations behave as perturbations on top of the $p=1$ truncation.

\subsection{Interpretation and partial deconfinement}
The original motivation for the search for instanton instabilities was to find evidence for the dual of the partially deconfined phase in the BPS phase diagram, which was paradoxically absent in previous analysis. Two consistent pictures of the partially deconfined phase emerge from our findings. 

In the first, the new phase indicated by the instability is identified with the partially deconfined phase. 
In this case, the `instanton' contributions could be identified with actual topological objects in the field theory. The condensation of topological objects are expected to coincide with the transition to partial deconfinement, as explained in Sec.~\ref{sec:pdec-instantons}. 
This picture also coheres well with the expectation that the transition from complete to partial deconfinement is described by a true, sharp phase transition, not merely a crossover, since the free energy potentials are not smooth, and a new condensate is formed. 
The emergence of a multi-cut saddle naturally creates a separation between degrees of freedom in colour space, as expected in the partially deconfined phase. Such detailed internal colour dynamics are seldom seen in examples other than the partially deconfined phase, with the partially Higgsed phase being the only close analogy.
Furthermore, the instanton condensation is driven by the statistical enhancement effect of maximising the gauge orbit (represented in the matrix model by the Vandermonde determinant), the same mechanism behind the transition to partial deconfinement presented in \cite{Hanada:2020uvt}.
Therefore, the proposal identifying the new phase as the partially deconfined phase creates a nice synthesis of all these expected signatures of partial deconfinement. 

In the second interpretation, the partially deconfined phase starts at the cusp and sits along the entire small black hole saddle. This is more consistent with the picture presented in \cite{Hanada:2018zxn,Hanada:2019czd,Hanada:2019rzv,Hanada:2022wcq}, where it is argued that the negative specific heat finds a natural explanation in the properties of partial deconfinement. Computations on the BFSS and BMN matrix models have been shown to follow this pattern, with the entire negative heat capacity saddle identified with the partially deconfined phase \cite{Bergner:2021goh}. 
In this case, the new instability should likely be identified with a transition to yet another phase, likely a different kind of partial deconfinement. Although this might seem less parsimonious, situations involving multiple classes of partially deconfined phases are known. In \cite{Hanada:2025rca}, it is shown that two kinds of partially deconfined phase, labelled PD-1 and PD-2, exist in QCD, distinguished by whether baryons have condensed. 

However, in this scenario, there is no evidence for the kind of dramatic phase transition from partial to complete deconfinement that we would expect. Indeed, the newly appreciated role of instanton condensation in marking the onset of the partially deconfined phase further raises expectations for a rigorous physical signature. Yet, the passage from large to small black hole appears to be completely smooth from the black hole perspective, with all thermodynamic potentials being smooth. It is not obvious how any nonanalyticity could emerge from this bulk perspective. The work in this paper might therefore be seen to challenge the expectations presented in \cite{Hanada:2018zxn,Hanada:2019czd,Hanada:2019rzv,Hanada:2022wcq}.

Finally, we comment on the puzzle of the location of the GWW transition. The GWW transition is found in the very small black hole limit, near the black hole/string transition region \cite{choi2024yang}. Before the insights of \cite{Hanada:2025rca}, briefly reviewed in Sec.~\ref{sec:nontrivial-gauge-condensates-GWW}, it was thought that this would imply that the transition to the partially deconfined phase takes place in this region, at entropy below $O(N^2)$, inconsistent with physical reasoning. Now, a more compelling explanation is that the finite chemical potential induces objects with non-trivial gauge orbits to condense at this point. Indeed, this is the point at which the charge of the black hole saddle first becomes nonzero at order $O(N^2)$, suggesting a corresponding condensate. If the objects have non-trivial gauge orbits, their condensation would explicitly break the gauge orbit and induce the GWW transition. We do not try to identify the condensing objects, but one candidate would be some kind of fuzzy sphere, dual to subdeterminant operators in the field theory. 

\subsection{Candidates for the new black hole phase}

In addition to being of intrinsic interest, understanding the new gravitational phase could help us to discern between the two interpretations of the partial deconfinement dual given above. We will focus here on two especially interesting candidates. The first is black hole configurations that are localised on the $S^5$ after a Gregory-Laflamme (GL) transition. The second is D3 brane nucleation. 

First, we discuss the possibility of a GL transition \cite{Gregory:1993vy, Gregory:1994bj}. In the thermal setting, there are strong arguments to associate black holes localised on the compact $S^5$ with partial deconfinement \cite{Hanada:2016pwv,Berenstein:2018lrm}. Analogy would imply that BPS black holes that are localised on the compact direction would also represent partially deconfined phases. The location of the instability is consistent with this possibility. Just as with the thermal GL instability, the BPS black hole instability sits just below the cusp. This is expected for transitions towards configurations that localise on the $S^5$. The distinction between large and small black holes is controlled by the $AdS$ scale. Meanwhile, the GL instability is dependent on the $AdS$ scale and the radius of the $S^5$, which in the IIB supergravity solution is equal to the $AdS$ scale. Therefore, we expect any GL transition to occur on a similar scale to the crossover between large and small black holes.

However, a GL transition in the BPS phase diagram is widely believed to be forbidden based on entropy scaling arguments advanced in \cite{Bhattacharyya:2010yg}. There, it was assumed that the endpoint of any hypothesised GL instability in the small charge limit is a black hole highly localised on the $S^5$, such that it can be approximated as a black hole in 10D flat space. Then, the BPS condition requires the spot-like black hole to spin at very high speeds around the compact $S^5$, setting an upper bound on the entropy that implies it is always subdominant to the standard black hole solution. However, it is possible that other endpoints exist with higher entropy.
For example, the standard thermal black hole is found to be unstable in the microcanonical ensemble not just against spot-like localisations, but also against localisation on the $S^5$ into belt configurations \cite{Dias:2015pda,Dias:2016eto}. Moreover, it is possible that we could have a localised hairy solution. An exhaustive search of hairy BPS black holes in $AdS_5 \times S^5$ at equal fugacity (the setup of the current work) was conducted in \cite{Dias:2024edd} within a consistent truncation of Type IIB supergravity that truncates modes of the compact $S^5$, and which further restricts the Cartan gauge fields $A_i$ associated with the $S^5$ isometries to $A_1=A_2=A_3$. Solutions localised on the $S^5$, having nontrivial tructure over the $S^5$, lie outside this truncation. It could be, therefore, that a GL transition takes us to hairy black holes that are localised on the $S^5$ in some way. This is consistent with the all the findings cited, and would be a good candidate for the partially deconfined phase. 

Next, we discuss the possibility of D3 brane nucleation. In \cite{chen2024giantgravitonexpansioneigenvalue}, a matrix model similar to the $p=1$ truncation was taken in the uniform eigenvalue distribution background, corresponding to the AdS vacuum in the bulk. Tunnelling eigenvalue contributions around this background were identified with terms in the giant graviton expansion, suggesting that these instantons correspond to D3 branes wrapped as giant gravitons. This eigenvalue tunnelling was exactly as described in Sec.~\ref{sec:matrix-model-detail-gww} for the ungapped distribution. The analogy to our situation, with gapped background distribution, is close. In the AdS vaccuum, these nonperturbative contributions are always suppressed, whereas in our case the instantons become dominant and begin to condense. If instantons in our setup admit an interpretation in terms of wrapped D3 brane configurations, this suggests that nucleation or condensation of these D3 branes is taking place. This would also fit well with a holographic description of partial deconfinement. D3 branes separating from the main stack could naturally describe internal colour dynamics and colour degrees of freedom leaving the deconfined phase.

D3 brane nucleation is a known mechanism by which black holes can become unstable and develop into interesting new phases, especially in the presence of chemical potentials \cite{Yamada:2008em, Henriksson:2019zph, Choi:2024xnv}. The endpoint of these instabilities could be hairy or multi-centre black holes. It might even describe the process of becoming localised on the $S^5$, thus merging this explanation with the Gregory-Laflamme transition proposed above. More common is nucleation to giant graviton solutions, wrapping an $S^3$ inside the compact $S^5$, on top of the black hole background. Another possibility is the nucleation of dual giant gravitons. This situation is similar to that discussed in \cite{Henriksson:2019zph}, which found that D3 branes could be ejected from rotating black holes and wrap an $S^3$ in the $AdS_5$. The resulting dual phase was interpreted as partially Higgsed. Since $N=4$ SYM is self-S dual, and S-duality maps the Higgs phase to confinement, we might expect this to naturally describe a partially deconfined phase, too. 

It is also possible that other kinds of hairy or multi-centre black holes are the correct description, as hypothesised in \cite{Benini:2018ywd}. It would be very interesting to conduct further work to resolve this question.

\subsection{Open questions and future directions}

If the new instability can indeed be with the onset of partial deconfinement, it offers a promising new route for uncovering properties of the partially deconfined phase and the mechanism of confinement. Indeed, the importance of the role of instantons in the partially deconfined phase \cite{Hanada:2023krw, Hanada:2023rlk} and the refined interpretation of the GWW point at finite chemical potential \cite{Hanada:2025rca} were both stimulated in part by the current work. The new gravity dual could be used to extract the physics of the partially deconfined phase at strong coupling, a topic currently wrapped in conjectures.
Understanding the instantons and their physical role seems especially exciting. Condensation of instantons is implicated in chiral symmetry breaking \cite{Diakonov:1995ea, Schafer:1996wv}, which is another expected signature of the partially deconfined phase \cite{Hanada:2018zxn,Hanada:2022wcq, Hanada:2021ksu}. It is also possible that these objects could play a role in the mechanism of confinement, such as in the fractional instanton liquid model \cite{Gonzalez-Arroyo:2023kqv, Soler:2025vwc}. 

Along with \cite{Hanada:2025rca}, the work presented here is one of the first explorations of the partial deconfinement at finite chemical potential. The finite chemical potential regime of QCD is important for describing the cores of neutron stars. Advancing the current work might lead us to important results such as the potential existence of the partially deconfined phase in these cores or the physical implications of a phase transition to the partially deconfined phase in the history of these stars.

It remains possible that the new phase does not describe partial deconfinement. This would also be very interesting. Like partial deconfinement, the phase seems to involve detailed internal colour dynamics, as shown by the colour eigenvalue density separating into two contours and the corresponding colour eigenvalues behaving distinctly. That such phases exist and are consistent with gauge invariance was not well-appreciated until partial deconfinement was proposed. Therefore, the field theory dual is likely very interesting and perhaps even exotic, even if it does not describe partial deconfinement. 

It is hoped that the work presented here could lead to a better understanding of how colour degrees of freedom are encoded under the holographic map. It would be especially exciting if the tunnelling eigenvalues can indeed be identified with nucleating D3 branes. Then, the paradigm by which colour eigenvalues can thought of as describing individual D3 branes can be extended. A simple identification of eigenvalues with D3 branes was widely believed to be untenable following arguments presented in e.g. \cite{Susskind:1998vk,Polchinski:1999br,Heemskerk:2012mn,Maldacena:2018vsr,Iizuka:2001cw}. However, a mistaken assumption in these arguments was recently recognised in \cite{Hanada:2021ipb}, inviting a reconsideration of this interpretation. 

Finally, it would be interesting to understand the instanton condensation instability through other approaches to calculating the superconformal index, such as the Bethe ansatz approach \cite{Benini:2018ywd, Aharony:2021zkr}. Finding a signature of the instability in the Bethe ansatz could open up new routes to understanding the meaning of the instanton condensation, for example by using methods similar to those in \cite{Aharony:2021zkr} to check for D3 brane nucleation. An instability in sufficiently small black holes was speculated in \cite{Benini:2018ywd} based on the onset of Stokes competition of the saddles found via the Bethe ansatz. However, this does not seem to match with the primary instability proposed in the current work, the $\pi$-type instanton condensation along the small black hole saddle, as it is located at much larger charge, before even the Hawking-Page transition. More plausibly, it could be related to the potential $d$-type instability exhibited by the $p=3$ truncated model and presented in Sec.~\ref{sec:results-p3}, although this suggestion is very speculative.

\section*{Acknowledgements}
I am very grateful to Pavel Buividovich, Vaibhav Gautam, Masanori Hanada, Seok Kim, Sameer Murthy, and Andy O'Bannon for useful discussions. I am especially grateful to Masanori Hanada and Andy O'Bannon for their constant encouragement and long patience throughout this project. The author would like to thank the Isaac Newton Institute for Mathematical Sciences, Cambridge, for support and hospitality during the programme ``Quantum Field Theory with Boundaries, Impurities, and Defects'' where work on this paper was undertaken. This work was supported by EPSRC grant no EP/R014604/1.

\appendix

\section{Higher truncation computations} \label{appendix:higher-truncations}
Here, we give some expressions for our calculations of the $\pi$-type instanton actions and pinching points for $p=2$ and $p=3$. Those for $p=1$ are already well-known and are given in Sec.~\ref{sec:matrix-model-detail-gww}.

Recall from eq.~\eqref{eq:y-general} that the spectral curve at truncation level $p$ for symmetric single-cut distributions can be expressed by,
\begin{equation}
    y(z) = \frac{g_p}{z^{p+1}}(z+1) \sqrt{(z-a)(z-a^{-1})} \prod_{i=1}^{p-1} (z-d_i)(z-d_i^{-1}),
\end{equation}
again using the definition $g_n=a_n \rho_n$, with $a_n(x)$ and the moments $\rho_n$ as defined in \eqref{eq:def-an-rho}. Following Appendix A in \cite{choi2024yang}, we are inspired to rewrite this as,
\begin{align} \label{eq:yz-as-Q}
    & y(z) = \nonumber
        \\
        & g_p\sqrt{z(z-a)(z-a^{-1})} \left(z^{p-\frac 1 2} + z^{-p-\frac 1 2} +\sum_{i=1}^{p-1} \frac{Q_i}{2 g_p} \left( z^{i-\frac 1 2} + z^{i+\frac 1 2} \right) \right)
\end{align}
It is shown in \cite{choi2024yang} that solving the saddle point (Dyson-Schwinger) equations allows the variables $B_i$ to be given in terms of the Legendre polynomials $P(x)$ by,
\begin{eqnarray}
    B_i = 2\sum_{l=0}^{p-i} g_{l+i} P_l \left(\frac{a + a^{-1}}{2} \right)
\end{eqnarray}
The roots of \eqref{eq:yz-as-Q} correspond to the endpoints $a$ and $a^{-1}$, the pinching point at $z=-1$, and the nontrivial pinching points $d_i$, which will be expressed in terms of the endpoints $a$ and $a^{-1}$. The endpoints $a$ can be found at any point along the black hole saddle according to the procedure described in Sec.~\ref{sec:index}. In this way, we can find the locations of the pinching points in the complex plane anywhere along the black hole saddles.

Finally, employing the Coulomb gas approach to matrix model calculations, we can integrate the spectral curve  as in \eqref{eq:Veff-integral} to obtain the instanton action corresponding to an eigenvalue tunneling to each pinching point, or, if symmetry demands, to each pair of pinching points. 

Below, we give expressions for the nontrivial pinching points $d_i$ for $p=2$ and $p=3$. We also give the instanton action for the $\pi$-type instantons. We do not provide here the actions we found for the $d$-type instantons, but they can be obtained according to the procedure mentioned above.

\subsection*{p=2}
The action $A_{2,\pi}$ for the $\pi$-type instanton actions at $p=2$ can be expressed as,
\begin{equation}
\begin{aligned}
    W &= (a-1)^2 \left((a+1)^2 g_2+2 a B_1\right) \left(\tanh
   ^{-1}\left(\sqrt{a}\right)+\coth ^{-1}\left(\sqrt{a}\right)\right) \\
    A_{2,\pi} &= \frac{\sqrt{a+\frac{1}{a}+2} \left(2 \sqrt{a} (a+1) \left(2 a B_1-(a+1)^2
   g_2\right)+ W \right) }{4 a^{3/2}
   (a+1)}
\end{aligned}
\end{equation}
The locations of the $d$-type pinching points can be expressed as,
\begin{equation}
    d_1^{\pm 1}=\frac{2 g_2-B_1 \pm \sqrt{-12 g_2^2-4 g_2 B_1+B_1^2}}{4 g_2}
\end{equation}

\subsection*{p=3}

The action $A_{3,\pi}$ for the $\pi$-type instanton actions at $p=3$ can be expressed as,
\begin{equation}
\begin{aligned}
W_1 &= 3 (a-1) a \left((a-1) \tanh ^{-1}\left(\sqrt{a}\right)-i
   \sqrt{-(a-1)^2} \text{csch}^{-1}\left(\sqrt{a-1}\right)\right) \\
  W_2 &= \left(\left(a^2+1\right) (a+1)^2 g_3+a \left(4 a B_1+(a+1)^2
   B_2\right)\right) \\
 A_{3,\pi} & = \frac{(a+1) \left(-2 (a+1)^2 a^{3/2} ((a (3 a-8)+3) g_3+3 a B_2)+24
   a^{7/2} B_1\right) + W_1 W_2}{24 a^4}
\end{aligned}
\end{equation}
The locations of the $d$-type pinching points can be expressed as,
\begin{equation}
\begin{aligned}
    v_1 &= \sqrt{
        B_2^2
        + g_3 (4 B_2 - 8 B_1)
        + 20 g_3^2
    } \\
    v_2 &= \sqrt{
        -40
        - \frac{4(2 B_1 + v_1)}{g_3}
        + \frac{2 B_2 (B_2 + v_1)}{g_3^2}} \\
    d_1^{\pm 1} &=  - \frac{1}{8g_3} (B_2 + v_1 + g_3 (-2 \pm v_2)) \\
    d_2^{\pm 1} &=  - \frac{1}{8g_3} (B_2 - v_1 + g_3 (-2 \pm v_2))
\end{aligned}
\end{equation}

\bibliographystyle{JHEP}
\bibliography{ref}

@article{Hanada:2019rzv,
    author = "Hanada, Masanori and Ishiki, Goro and Watanabe, Hiromasa",
    title = "{Partial deconfinement in gauge theories}",
    eprint = "1911.11465",
    archivePrefix = "arXiv",
    primaryClass = "hep-lat",
    reportNumber = "UTHEP-739",
    doi = "10.22323/1.363.0055",
    journal = "PoS",
    volume = "LATTICE2019",
    pages = "055",
    year = "2019"
}

@article{Hanada:2021ksu,
    author = "Hanada, Masanori and Holden, Jack and Knaggs, Matthew and O'Bannon, Andy",
    title = "{Global symmetries and partial confinement}",
    eprint = "2112.11398",
    archivePrefix = "arXiv",
    primaryClass = "hep-th",
    reportNumber = "DMUS-MP-21/11",
    doi = "10.1007/JHEP03(2022)118",
    journal = "JHEP",
    volume = "03",
    pages = "118",
    year = "2022"
}

@article{Hanada:2019kue,
    author = "Hanada, Masanori and Robinson, Brandon",
    title = "{Partial-Symmetry-Breaking Phase Transitions}",
    eprint = "1911.06223",
    archivePrefix = "arXiv",
    primaryClass = "hep-th",
    doi = "10.1103/PhysRevD.102.096013",
    journal = "Phys. Rev. D",
    volume = "102",
    number = "9",
    pages = "096013",
    year = "2020"
}

@article{Hanada:2018zxn,
    author = "Hanada, Masanori and Ishiki, Goro and Watanabe, Hiromasa",
    title = "{Partial Deconfinement}",
    eprint = "1812.05494",
    archivePrefix = "arXiv",
    primaryClass = "hep-th",
    reportNumber = "UTHEP-728",
    doi = "10.1007/JHEP03(2019)145",
    journal = "JHEP",
    volume = "03",
    pages = "145",
    year = "2019",
    note = "[Erratum: JHEP 10, 029 (2019)]"
}

@article{Hanada:2016pwv,
    author = "Hanada, Masanori and Maltz, Jonathan",
    title = "{A proposal of the gauge theory description of the small Schwarzschild black hole in AdS$_5\times$S$^5$}",
    eprint = "1608.03276",
    archivePrefix = "arXiv",
    primaryClass = "hep-th",
    reportNumber = "SU-ITP-16-14, YITP-16-95",
    doi = "10.1007/JHEP02(2017)012",
    journal = "JHEP",
    volume = "02",
    pages = "012",
    year = "2017"
}

@article{Berenstein:2018lrm,
    author = "Berenstein, David",
    title = "{Submatrix deconfinement and small black holes in AdS}",
    eprint = "1806.05729",
    archivePrefix = "arXiv",
    primaryClass = "hep-th",
    doi = "10.1007/JHEP09(2018)054",
    journal = "JHEP",
    volume = "09",
    pages = "054",
    year = "2018"
}

@article{Hanada:2019czd,
    author = "Hanada, Masanori and Jevicki, Antal and Peng, Cheng and Wintergerst, Nico",
    title = "{Anatomy of Deconfinement}",
    eprint = "1909.09118",
    archivePrefix = "arXiv",
    primaryClass = "hep-th",
    doi = "10.1007/JHEP12(2019)167",
    journal = "JHEP",
    volume = "12",
    pages = "167",
    year = "2019"
}

@article{Hanada:2020uvt,
    author = "Hanada, Masanori and Shimada, Hidehiko and Wintergerst, Nico",
    title = "{Color confinement and Bose-Einstein condensation}",
    eprint = "2001.10459",
    archivePrefix = "arXiv",
    primaryClass = "hep-th",
    doi = "10.1007/JHEP08(2021)039",
    journal = "JHEP",
    volume = "08",
    pages = "039",
    year = "2021"
}

@article{Hanada:2023krw,
    author = "Hanada, Masanori and Ohata, Hiroki and Shimada, Hidehiko and Watanabe, Hiromasa",
    title = "{A New Perspective on Thermal Transition in QCD}",
    eprint = "2310.01940",
    archivePrefix = "arXiv",
    primaryClass = "hep-th",
    reportNumber = "YITP-23-121",
    doi = "10.1093/ptep/ptae044",
    journal = "PTEP",
    volume = "2024",
    number = "4",
    pages = "041B02",
    year = "2024"
}

@article{Hanada:2023rlk,
    author = "Hanada, Masanori and Watanabe, Hiromasa",
    title = "{On Thermal Transition in QCD}",
    eprint = "2310.07533",
    archivePrefix = "arXiv",
    primaryClass = "hep-th",
    reportNumber = "YITP-23-125",
    doi = "10.1093/ptep/ptae033",
    journal = "PTEP",
    volume = "2024",
    number = "4",
    pages = "043B02",
    year = "2024"
}

@article{Aharony:2003sx,
    author = "Aharony, Ofer and Marsano, Joseph and Minwalla, Shiraz and Papadodimas, Kyriakos and Van Raamsdonk, Mark",
    editor = "Doebner, H. D. and Dobrev, V. K.",
    title = "{The Hagedorn - deconfinement phase transition in weakly coupled large N gauge theories}",
    eprint = "hep-th/0310285",
    archivePrefix = "arXiv",
    reportNumber = "WIS-29-03-DPP",
    doi = "10.4310/ATMP.2004.v8.n4.a1",
    journal = "Adv. Theor. Math. Phys.",
    volume = "8",
    pages = "603--696",
    year = "2004"
}

@article{Witten:1998zw,
    author = "Witten, Edward",
    editor = "Bergstrom, L. and Lindstrom, U.",
    title = "{Anti-de Sitter space, thermal phase transition, and confinement in gauge theories}",
    eprint = "hep-th/9803131",
    archivePrefix = "arXiv",
    reportNumber = "IASSNS-HEP-98-21",
    doi = "10.4310/ATMP.1998.v2.n3.a3",
    journal = "Adv. Theor. Math. Phys.",
    volume = "2",
    pages = "505--532",
    year = "1998"
}

@article{Sundborg:1999ue,
    author = "Sundborg, Bo",
    title = "{The Hagedorn transition, deconfinement and N=4 SYM theory}",
    eprint = "hep-th/9908001",
    archivePrefix = "arXiv",
    reportNumber = "USITP-99-06",
    doi = "10.1016/S0550-3213(00)00044-4",
    journal = "Nucl. Phys. B",
    volume = "573",
    pages = "349--363",
    year = "2000"
}

@article{R_melsberger_2006,
   title={Counting chiral primaries in , superconformal field theories},
   volume={747},
   ISSN={0550-3213},
   url={http://dx.doi.org/10.1016/j.nuclphysb.2006.03.037},
   DOI={10.1016/j.nuclphysb.2006.03.037},
   number={3},
   journal={Nuclear Physics B},
   publisher={Elsevier BV},
   author={Römelsberger, Christian},
   year={2006},
   month=jul, pages={329–353} 
}

@article{Kinney_2007,
   title={An Index for 4 Dimensional Super Conformal Theories},
   volume={275},
   ISSN={1432-0916},
   url={http://dx.doi.org/10.1007/s00220-007-0258-7},
   DOI={10.1007/s00220-007-0258-7},
   number={1},
   journal={Communications in Mathematical Physics},
   publisher={Springer Science and Business Media LLC},
   author={Kinney, Justin and Maldacena, Juan and Minwalla, Shiraz and Raju, Suvrat},
   year={2007},
   month=jun, pages={209–254} 
}

@article{Ezroura_2022,
   title={The phase diagram of BPS black holes in AdS5},
   volume={2022},
   ISSN={1029-8479},
   url={http://dx.doi.org/10.1007/JHEP09(2022)033},
   DOI={10.1007/jhep09(2022)033},
   number={9},
   journal={Journal of High Energy Physics},
   publisher={Springer Science and Business Media LLC},
   author={Ezroura, Nizar and Larsen, Finn and Liu, Zhihan and Zeng, Yangwenxiao},
   year={2022},
   month=sep 
}

@article{choi2024yang,
  title={The Yang-Mills duals of small AdS black holes},
  author={Choi, Sunjin and Jeong, Saebyeok and Kim, Seok},
  journal={Journal of High Energy Physics},
  volume={2024},
  number={7},
  pages={1--60},
  year={2024},
  publisher={Springer}
}

@misc{chen2024giantgravitonexpansioneigenvalue,
      title={Giant graviton expansion from eigenvalue instantons}, 
      author={Yiming Chen and Raghu Mahajan and Haifeng Tang},
      year={2024},
      eprint={2407.08155},
      archivePrefix={arXiv},
      primaryClass={hep-th},
      url={https://arxiv.org/abs/2407.08155}, 
}

@article{Cabo_Bizet_2019,
   title={Microscopic origin of the Bekenstein-Hawking entropy of supersymmetric AdS5 black holes},
   volume={2019},
   ISSN={1029-8479},
   url={http://dx.doi.org/10.1007/JHEP10(2019)062},
   DOI={10.1007/jhep10(2019)062},
   number={10},
   journal={Journal of High Energy Physics},
   publisher={Springer Science and Business Media LLC},
   author={Cabo-Bizet, Alejandro and Cassani, Davide and Martelli, Dario and Murthy, Sameer},
   year={2019},
   month=oct }

@misc{choi2024largeadsblackholes,
      title={Large AdS black holes from QFT}, 
      author={Sunjin Choi and Joonho Kim and Seok Kim and June Nahmgoong},
      year={2024},
      eprint={1810.12067},
      archivePrefix={arXiv},
      primaryClass={hep-th},
      url={https://arxiv.org/abs/1810.12067}, 
}

@article{Benini:2018ywd,
    author = "Benini, Francesco and Milan, Elisa",
    title = "{Black Holes in 4D $\mathcal{N}$=4 Super-Yang-Mills Field Theory}",
    eprint = "1812.09613",
    archivePrefix = "arXiv",
    primaryClass = "hep-th",
    reportNumber = "SISSA 56/2018/FISI",
    doi = "10.1103/PhysRevX.10.021037",
    journal = "Phys. Rev. X",
    volume = "10",
    number = "2",
    pages = "021037",
    year = "2020"
}

@article{Gross:1980he,
      author         = "Gross, D. J. and Witten, Edward",
      title          = "{Possible Third Order Phase Transition in the Large N
                        Lattice Gauge Theory}",
      journal        = "Phys. Rev.",
      volume         = "D21",
      year           = "1980",
      pages          = "446-453",
      doi            = "10.1103/PhysRevD.21.446",
      SLACcitation   = "%%CITATION = PHRVA,D21,446;%%"
}

@article{Wadia:2012fr,
      author         = "Wadia, Spenta R.",
      title          = "{A Study of U(N) Lattice Gauge Theory in 2-dimensions}",
      year           = "2012",
      eprint         = "1212.2906",
      archivePrefix  = "arXiv",
      primaryClass   = "hep-th",
      reportNumber   = "ICTS-2012-13, TIFR-TH-2012-47",
      SLACcitation   = "%%CITATION = ARXIV:1212.2906;%%"
}

@article{Maldacena_1999,
   title={The Large-N Limit of Superconformal Field Theories and Supergravity},
   volume={38},
   ISSN={1572-9575},
   url={http://dx.doi.org/10.1023/A:1026654312961},
   DOI={10.1023/a:1026654312961},
   number={4},
   journal={International Journal of Theoretical Physics},
   publisher={Springer Science and Business Media LLC},
   author={Maldacena, Juan},
   year={1999},
   month=apr, pages={1113–1133} }

@article{Marino:2008ya,
    author = "Marino, Marcos",
    title = "{Nonperturbative effects and nonperturbative definitions in matrix models and topological strings}",
    eprint = "0805.3033",
    archivePrefix = "arXiv",
    primaryClass = "hep-th",
    doi = "10.1088/1126-6708/2008/12/114",
    journal = "JHEP",
    volume = "12",
    pages = "114",
    year = "2008"
}

@article{Neuberger:1980qh,
    author = "Neuberger, H.",
    title = "{Nonperturbative Contributions in Models With a Nonanalytic Behavior at Infinite $N$}",
    reportNumber = "UCB-PTH-80/9",
    doi = "10.1016/0550-3213(81)90238-8",
    journal = "Nucl. Phys. B",
    volume = "179",
    pages = "253--282",
    year = "1981"
}

@article{Neuberger:1980as,
    author = "Neuberger, H.",
    title = "{INSTANTONS AS A BRIDGEHEAD AT N = infinity}",
    reportNumber = "UCB-PTH-80/5",
    doi = "10.1016/0370-2693(80)90858-8",
    journal = "Phys. Lett. B",
    volume = "94",
    pages = "199--202",
    year = "1980"
}

@article{Buividovich:2015oju,
    author = "Buividovich, P. V. and Dunne, Gerald V. and Valgushev, S. N.",
    title = "{Complex Path Integrals and Saddles in Two-Dimensional Gauge Theory}",
    eprint = "1512.09021",
    archivePrefix = "arXiv",
    primaryClass = "hep-th",
    doi = "10.1103/PhysRevLett.116.132001",
    journal = "Phys. Rev. Lett.",
    volume = "116",
    number = "13",
    pages = "132001",
    year = "2016"
}

@article{Alvarez:2016rmo,
    author = "{\'A}lvarez, Gabriel and Mart{\'\i}nez Alonso, Luis and Medina, Elena",
    title = "{Complex saddles in the Gross-Witten-Wadia matrix model}",
    eprint = "1610.09948",
    archivePrefix = "arXiv",
    primaryClass = "hep-th",
    doi = "10.1103/PhysRevD.94.105010",
    journal = "Phys. Rev. D",
    volume = "94",
    number = "10",
    pages = "105010",
    year = "2016"
}

@article{Fujimori:2021oqg,
    author = "Fujimori, Toshiaki and Honda, Masazumi and Kamata, Syo and Misumi, Tatsuhiro and Sakai, Norisuke and Yoda, Takuya",
    title = "{Quantum phase transition and resurgence: Lessons from three-dimensional $\mathcal{N}=4$ supersymmetric quantum electrodynamics}",
    eprint = "2103.13654",
    archivePrefix = "arXiv",
    primaryClass = "hep-th",
    reportNumber = "YITP-21-13, KUNS-2859",
    doi = "10.1093/ptep/ptab086",
    journal = "PTEP",
    volume = "2021",
    number = "10",
    pages = "103B04",
    year = "2021"
}

@article{Dias:2024edd,
    author = "Dias, Oscar J. C. and Mitra, Prahar and Santos, Jorge E.",
    title = "{Charged rotating hairy black holes in AdS$_{5}$ {\texttimes} S$^{5}$: unveiling their secrets}",
    eprint = "2411.18712",
    archivePrefix = "arXiv",
    primaryClass = "hep-th",
    doi = "10.1007/JHEP06(2025)051",
    journal = "JHEP",
    volume = "06",
    pages = "051",
    year = "2025"
}

@article{Dias:2022eyq,
    author = "Dias, Oscar J. C. and Mitra, Prahar and Santos, Jorge E.",
    title = "{New phases of $ \mathcal{N} $ = 4 SYM at finite chemical potential}",
    eprint = "2207.07134",
    archivePrefix = "arXiv",
    primaryClass = "hep-th",
    doi = "10.1007/JHEP05(2023)053",
    journal = "JHEP",
    volume = "05",
    pages = "053",
    year = "2023"
}

@article{Dias:2016eto,
    author = "Dias, {\'O}scar J. C. and Santos, Jorge E. and Way, Benson",
    title = "{Localised $AdS_5\times S^5$ Black Holes}",
    eprint = "1605.04911",
    archivePrefix = "arXiv",
    primaryClass = "hep-th",
    doi = "10.1103/PhysRevLett.117.151101",
    journal = "Phys. Rev. Lett.",
    volume = "117",
    number = "15",
    pages = "151101",
    year = "2016"
}

@article{Dias:2015pda,
    author = "Dias, {\'O}scar J. C. and Santos, Jorge E. and Way, Benson",
    title = "{Lumpy AdS$_{5}${\texttimes} S$^{5}$ black holes and black belts}",
    eprint = "1501.06574",
    archivePrefix = "arXiv",
    primaryClass = "hep-th",
    doi = "10.1007/JHEP04(2015)060",
    journal = "JHEP",
    volume = "04",
    pages = "060",
    year = "2015"
}

@article{Choi:2024xnv,
    author = "Choi, Sunjin and Jain, Diksha and Kim, Seok and Krishna, Vineeth and Lee, Eunwoo and Minwalla, Shiraz and Patel, Chintan",
    title = "{Dual dressed black holes as the end point of the charged superradiant instability in $\mathcal{N} = 4$ Yang Mills}",
    eprint = "2409.18178",
    archivePrefix = "arXiv",
    primaryClass = "hep-th",
    reportNumber = "TIFR/TH/24-19, LCTP-24-17",
    doi = "10.21468/SciPostPhys.18.4.137",
    journal = "SciPost Phys.",
    volume = "18",
    number = "4",
    pages = "137",
    year = "2025"
}

@article{Choi:2021rxi,
    author = "Choi, Sunjin and Jeong, Saebyeok and Kim, Seok and Lee, Eunwoo",
    title = "{Exact QFT duals of AdS black holes}",
    eprint = "2111.10720",
    archivePrefix = "arXiv",
    primaryClass = "hep-th",
    reportNumber = "KIAS-P21054, SNUTP21-002",
    doi = "10.1007/JHEP09(2023)138",
    journal = "JHEP",
    volume = "09",
    pages = "138",
    year = "2023"
}

@article{Marino:2008vx,
    author = "Marino, Marcos and Schiappa, Ricardo and Weiss, Marlene",
    title = "{Multi-Instantons and Multi-Cuts}",
    eprint = "0809.2619",
    archivePrefix = "arXiv",
    primaryClass = "hep-th",
    reportNumber = "CERN-PH-TH-2008-186",
    doi = "10.1063/1.3097755",
    journal = "J. Math. Phys.",
    volume = "50",
    pages = "052301",
    year = "2009"
}

@article{Marino:2022rpz,
    author = "Marino, Marcos and Schiappa, Ricardo and Schwick, Maximilian",
    title = "{New Instantons for Matrix Models}",
    eprint = "2210.13479",
    archivePrefix = "arXiv",
    primaryClass = "hep-th",
    month = "10",
    year = "2022"
}

@article{Gregory:1993vy,
    author = "Gregory, R. and Laflamme, R.",
    title = "{Black strings and p-branes are unstable}",
    eprint = "hep-th/9301052",
    archivePrefix = "arXiv",
    doi = "10.1103/PhysRevLett.70.2837",
    journal = "Phys. Rev. Lett.",
    volume = "70",
    pages = "2837--2840",
    year = "1993"
}

@article{Gregory:1994bj,
    author = "Gregory, Ruth and Laflamme, Raymond",
    title = "{The Instability of charged black strings and p-branes}",
    eprint = "hep-th/9404071",
    archivePrefix = "arXiv",
    reportNumber = "DAMTP-R-94-7, LA-UR-93-4473",
    doi = "10.1016/0550-3213(94)90206-2",
    journal = "Nucl. Phys. B",
    volume = "428",
    pages = "399--434",
    year = "1994"
}

@article{Copetti:2020dil,
    author = "Copetti, Christian and Grassi, Alba and Komargodski, Zohar and Tizzano, Luigi",
    title = "{Delayed deconfinement and the Hawking-Page transition}",
    eprint = "2008.04950",
    archivePrefix = "arXiv",
    primaryClass = "hep-th",
    doi = "10.1007/JHEP04(2022)132",
    journal = "JHEP",
    volume = "04",
    pages = "132",
    year = "2022"
}

@article{Bhattacharyya:2010yg,
    author = "Bhattacharyya, Sayantani and Minwalla, Shiraz and Papadodimas, Kyriakos",
    title = "{Small Hairy Black Holes in $AdS_5 x S^5$}",
    eprint = "1005.1287",
    archivePrefix = "arXiv",
    primaryClass = "hep-th",
    reportNumber = "TIFR-TH-ITFA-10-13",
    doi = "10.1007/JHEP11(2011)035",
    journal = "JHEP",
    volume = "11",
    pages = "035",
    year = "2011"
}

@article{Hanada:2021ipb,
    author = "Hanada, Masanori",
    title = "{Bulk geometry in gauge/gravity duality and color degrees of freedom}",
    eprint = "2102.08982",
    archivePrefix = "arXiv",
    primaryClass = "hep-th",
    reportNumber = "DMUS-MP-21/02",
    doi = "10.1103/PhysRevD.103.106007",
    journal = "Phys. Rev. D",
    volume = "103",
    number = "10",
    pages = "106007",
    year = "2021"
}

@article{Hanada:2025rca,
    author = "Hanada, Masanori and Holden, Jack and Watanabe, Hiromasa",
    title = "{New phases in QCD at finite temperature and chemical potential}",
    eprint = "2509.04671",
    archivePrefix = "arXiv",
    primaryClass = "hep-th",
    month = "9",
    year = "2025"
}

@article{Kolbig:1981qz,
    author = "Kolbig, K. S. and Ruhl, W.",
    title = "{COMPLEX ZEROS OF THE PARTITION FUNCTION FOR TWO-DIMENSIONAL U(N) LATTICE GAUGE THEORIES}",
    reportNumber = "CERN-TH-3125",
    doi = "10.1007/BF01548610",
    journal = "Z. Phys. C",
    volume = "12",
    pages = "135--143",
    year = "1982"
}

@article{Gladden:2024ssb,
    author = "Gladden, Liam and Ivo, Victor and Kovtun, Pavel and Starinets, Andrei O.",
    title = "{Instability in N=4 supersymmetric Yang-Mills theory at finite density}",
    eprint = "2412.12353",
    archivePrefix = "arXiv",
    primaryClass = "hep-th",
    doi = "10.1103/PhysRevD.111.086030",
    journal = "Phys. Rev. D",
    volume = "111",
    number = "8",
    pages = "086030",
    year = "2025"
}

@article{Cabo-Bizet:2019eaf,
    author = "Cabo-Bizet, Alejandro and Murthy, Sameer",
    title = "{Supersymmetric phases of 4d $ \mathcal{N} $ = 4 SYM at large $N$}",
    eprint = "1909.09597",
    archivePrefix = "arXiv",
    primaryClass = "hep-th",
    doi = "10.1007/JHEP09(2020)184",
    journal = "JHEP",
    volume = "09",
    pages = "184",
    year = "2020"
}

@article{Gautam:2022exf,
    author = "Gautam, Vaibhav and Hanada, Masanori and Holden, Jack and Rinaldi, Enrico",
    title = "{Linear confinement in the partially-deconfined phase}",
    eprint = "2208.14402",
    archivePrefix = "arXiv",
    primaryClass = "hep-th",
    reportNumber = "DMUS-MP-22/14, RIKEN-iTHEMS-Report-22",
    doi = "10.1007/JHEP03(2023)195",
    journal = "JHEP",
    volume = "03",
    pages = "195",
    year = "2023"
}

@article{Marino:2007te,
    author = "Marino, Marcos and Schiappa, Ricardo and Weiss, Marlene",
    title = "{Nonperturbative Effects and the Large-Order Behavior of Matrix Models and Topological Strings}",
    eprint = "0711.1954",
    archivePrefix = "arXiv",
    primaryClass = "hep-th",
    reportNumber = "CERN-PH-TH-2007-218",
    doi = "10.4310/CNTP.2008.v2.n2.a3",
    journal = "Commun. Num. Theor. Phys.",
    volume = "2",
    pages = "349--419",
    year = "2008"
}

@article{Dunne:2012ae,
    author = "Dunne, Gerald V. and Unsal, Mithat",
    title = "{Resurgence and Trans-series in Quantum Field Theory: The CP(N-1) Model}",
    eprint = "1210.2423",
    archivePrefix = "arXiv",
    primaryClass = "hep-th",
    doi = "10.1007/JHEP11(2012)170",
    journal = "JHEP",
    volume = "11",
    pages = "170",
    year = "2012"
}

@article{Hanada:2022wcq,
    author = "Hanada, Masanori and Watanabe, Hiromasa",
    title = "{Partial deconfinement: a brief overview}",
    eprint = "2210.11216",
    archivePrefix = "arXiv",
    primaryClass = "hep-th",
    reportNumber = "DMUS-MP-22/16, KEK-TH-2445, YITP-22-118",
    doi = "10.1140/epjs/s11734-022-00709-0",
    journal = "Eur. Phys. J. ST",
    volume = "232",
    number = "3",
    pages = "333--337",
    year = "2023"
}

@article{Henriksson:2019zph,
    author = "Henriksson, Oscar and Hoyos, Carlos and Jokela, Niko",
    title = "{Novel color superconducting phases of $\cal{N}$ = 4 super Yang-Mills at strong coupling}",
    eprint = "1907.01562",
    archivePrefix = "arXiv",
    primaryClass = "hep-th",
    reportNumber = "HIP-2019-19/TH",
    doi = "10.1007/JHEP09(2019)088",
    journal = "JHEP",
    volume = "09",
    pages = "088",
    year = "2019"
}

@article{Susskind:1998vk,
    author = "Susskind, Leonard",
    editor = "Burgess, C. P. and Myers, Robert C.",
    title = "{Holography in the flat space limit}",
    eprint = "hep-th/9901079",
    archivePrefix = "arXiv",
    doi = "10.1063/1.1301570",
    journal = "AIP Conf. Proc.",
    volume = "493",
    number = "1",
    pages = "98--112",
    year = "1999"
}

@article{Polchinski:1999br,
    author = "Polchinski, Joseph",
    editor = "Iso, S. and Kawai, H. and Natsuume, M.",
    title = "{M theory and the light cone}",
    eprint = "hep-th/9903165",
    archivePrefix = "arXiv",
    reportNumber = "NSF-ITP-99-17",
    doi = "10.1143/PTPS.134.158",
    journal = "Prog. Theor. Phys. Suppl.",
    volume = "134",
    pages = "158--170",
    year = "1999"
}

@article{Heemskerk:2012mn,
    author = "Heemskerk, Idse and Marolf, Donald and Polchinski, Joseph and Sully, James",
    title = "{Bulk and Transhorizon Measurements in AdS/CFT}",
    eprint = "1201.3664",
    archivePrefix = "arXiv",
    primaryClass = "hep-th",
    reportNumber = "NSF-KITP-12-081",
    doi = "10.1007/JHEP10(2012)165",
    journal = "JHEP",
    volume = "10",
    pages = "165",
    year = "2012"
}

@article{Maldacena:2018vsr,
    author = "Maldacena, Juan and Milekhin, Alexey",
    title = "{To gauge or not to gauge?}",
    eprint = "1802.00428",
    archivePrefix = "arXiv",
    primaryClass = "hep-th",
    doi = "10.1007/JHEP04(2018)084",
    journal = "JHEP",
    volume = "04",
    pages = "084",
    year = "2018"
}

@article{Iizuka:2001cw,
    author = "Iizuka, Norihiro and Kabat, Daniel N. and Lifschytz, Gilad and Lowe, David A.",
    title = "{Probing black holes in nonperturbative gauge theory}",
    eprint = "hep-th/0108006",
    archivePrefix = "arXiv",
    reportNumber = "BROWN-HET-1276, CU-TP-1025",
    doi = "10.1103/PhysRevD.65.024012",
    journal = "Phys. Rev. D",
    volume = "65",
    pages = "024012",
    year = "2002"
}

@article{Schnitzer:2004qt,
    author = "Schnitzer, Howard J.",
    title = "{Confinement/deconfinement transition of large N gauge theories with N(f) fundamentals: N(f)/N finite}",
    eprint = "hep-th/0402219",
    archivePrefix = "arXiv",
    reportNumber = "BRX-TH-534",
    doi = "10.1016/j.nuclphysb.2004.06.057",
    journal = "Nucl. Phys. B",
    volume = "695",
    pages = "267--282",
    year = "2004"
}

@article{Witten:1982df,
    author = "Witten, Edward",
    title = "{Constraints on Supersymmetry Breaking}",
    reportNumber = "PRINT-82-0163 (PRINCETON)",
    doi = "10.1016/0550-3213(82)90071-2",
    journal = "Nucl. Phys. B",
    volume = "202",
    pages = "253",
    year = "1982"
}

@article{Rossi:1996hs,
    author = "Rossi, Paolo and Campostrini, Massimo and Vicari, Ettore",
    title = "{The Large N expansion of unitary matrix models}",
    eprint = "hep-lat/9609003",
    archivePrefix = "arXiv",
    reportNumber = "IFUP-TH-54-96",
    doi = "10.1016/S0370-1573(98)00003-9",
    journal = "Phys. Rept.",
    volume = "302",
    pages = "143--209",
    year = "1998"
}

@article{Bars:1979xb,
    author = "Bars, Itzhak and Green, Frederic",
    title = "{Complete Integration of U ($N$) Lattice Gauge Theory in a Large $N$ Limit}",
    reportNumber = "Print-79-0416 (IAS,PRINCETON)",
    doi = "10.1103/PhysRevD.20.3311",
    journal = "Phys. Rev. D",
    volume = "20",
    pages = "3311",
    year = "1979"
}

@article{Goldschmidt:1979hq,
    author = "Goldschmidt, Yadin Y.",
    title = "{1/$N$ Expansion in Two-dimensional Lattice Gauge Theory}",
    reportNumber = "SACLAY-DPh-T 79/153",
    doi = "10.1063/1.524600",
    journal = "J. Math. Phys.",
    volume = "21",
    pages = "1842",
    year = "1980"
}

@article{Bergner:2021goh,
    author = {Bergner, Georg and Bodendorfer, Norbert and Hanada, Masanori and Pateloudis, Stratos and Rinaldi, Enrico and Sch{\"a}fer, Andreas and Vranas, Pavlos and Watanabe, Hiromasa},
    collaboration = "Monte Carlo String/M-theory (MCSMC), MCSMC",
    title = "{Confinement/deconfinement transition in the D0-brane matrix model {\textemdash} A signature of M-theory?}",
    eprint = "2110.01312",
    archivePrefix = "arXiv",
    primaryClass = "hep-th",
    reportNumber = "LLNL-JRNL-824792, RIKEN-iTHEMS-Report-21, UTHEP-759, DMUS-MP-21/13",
    doi = "10.1007/JHEP05(2022)096",
    journal = "JHEP",
    volume = "05",
    pages = "096",
    year = "2022"
}

@article{Witten:1998qj,
    author = "Witten, Edward",
    title = "{Anti de Sitter space and holography}",
    eprint = "hep-th/9802150",
    archivePrefix = "arXiv",
    reportNumber = "IASSNS-HEP-98-15",
    doi = "10.4310/ATMP.1998.v2.n2.a2",
    journal = "Adv. Theor. Math. Phys.",
    volume = "2",
    pages = "253--291",
    year = "1998"
}

@article{Gutowski:2004yv,
    author = "Gutowski, Jan B. and Reall, Harvey S.",
    title = "{General supersymmetric AdS(5) black holes}",
    eprint = "hep-th/0401129",
    archivePrefix = "arXiv",
    reportNumber = "NSF-KITP-04-10",
    doi = "10.1088/1126-6708/2004/04/048",
    journal = "JHEP",
    volume = "04",
    pages = "048",
    year = "2004"
}

@article{Gutowski:2004ez,
    author = "Gutowski, Jan B. and Reall, Harvey S.",
    title = "{Supersymmetric AdS(5) black holes}",
    eprint = "hep-th/0401042",
    archivePrefix = "arXiv",
    reportNumber = "NSF-KITP-04-02",
    doi = "10.1088/1126-6708/2004/02/006",
    journal = "JHEP",
    volume = "02",
    pages = "006",
    year = "2004"
}

@article{Diakonov:1995ea,
    author = "Diakonov, Dmitri",
    editor = "Di Giacomo, A. and Diakonov, Dmitri",
    title = "{Chiral symmetry breaking by instantons}",
    eprint = "hep-ph/9602375",
    archivePrefix = "arXiv",
    doi = "10.3254/978-1-61499-215-8-397",
    journal = "Proc. Int. Sch. Phys. Fermi",
    volume = "130",
    pages = "397--432",
    year = "1996"
}

@article{Schafer:1996wv,
    author = {Sch{\"a}fer, Thomas and Shuryak, Edward V.},
    title = "{Instantons in QCD}",
    eprint = "hep-ph/9610451",
    archivePrefix = "arXiv",
    reportNumber = "DOE-ER-40561-293, INT-96-00-150",
    doi = "10.1103/RevModPhys.70.323",
    journal = "Rev. Mod. Phys.",
    volume = "70",
    pages = "323--426",
    year = "1998"
}

@article{Yamada:2008em,
    author = "Yamada, Daiske",
    title = "{Fragmentation of Spinning Branes}",
    eprint = "0802.3508",
    archivePrefix = "arXiv",
    primaryClass = "hep-th",
    doi = "10.1088/0264-9381/25/14/145006",
    journal = "Class. Quant. Grav.",
    volume = "25",
    pages = "145006",
    year = "2008"
}

@article{Soler:2025vwc,
    author = "Soler, Ivan and Bergner, Georg and Gonzalez-Arroyo, Antonio",
    title = "{Fractional instantons and Confinement: first results on a $T_2 \times R^2$ roadmap}",
    eprint = "2502.09463",
    archivePrefix = "arXiv",
    primaryClass = "hep-lat",
    reportNumber = "IFT-UAM/CSIC-25-14",
    doi = "10.22323/1.466.0408",
    journal = "PoS",
    volume = "LATTICE2024",
    pages = "408",
    year = "2025"
}

@article{Gonzalez-Arroyo:2023kqv,
    author = "Gonzalez-Arroyo, Antonio",
    title = "{On the fractional instanton liquid picture of the Yang-Mills vacuum and Confinement}",
    eprint = "2302.12356",
    archivePrefix = "arXiv",
    primaryClass = "hep-th",
    reportNumber = "IFT-UAM/CSIC-23-20",
    month = "2",
    year = "2023"
}

@article{David:1990sk,
    author = "David, Francois",
    title = "{Phases of the large N matrix model and nonperturbative effects in 2-d gravity}",
    reportNumber = "SACLAY-SPH-T-90-090",
    doi = "10.1016/0550-3213(91)90202-9",
    journal = "Nucl. Phys. B",
    volume = "348",
    pages = "507--524",
    year = "1991"
}

@article{Aharony:2021zkr,
    author = "Aharony, Ofer and Benini, Francesco and Mamroud, Ohad and Milan, Elisa",
    title = "{A gravity interpretation for the Bethe Ansatz expansion of the $\mathcal{N}=4$ SYM index}",
    eprint = "2104.13932",
    archivePrefix = "arXiv",
    primaryClass = "hep-th",
    reportNumber = "SISSA 01/2021/FISI",
    doi = "10.1103/PhysRevD.104.086026",
    journal = "Phys. Rev. D",
    volume = "104",
    pages = "086026",
    year = "2021"
}

@article{Aharony:2005bq,
    author = "Aharony, Ofer and Marsano, Joseph and Minwalla, Shiraz and Papadodimas, Kyriakos and Van Raamsdonk, Mark",
    title = "{A First order deconfinement transition in large N Yang-Mills theory on a small S**3}",
    eprint = "hep-th/0502149",
    archivePrefix = "arXiv",
    reportNumber = "WIS-03-05-JAN-DPP",
    doi = "10.1103/PhysRevD.71.125018",
    journal = "Phys. Rev. D",
    volume = "71",
    pages = "125018",
    year = "2005"
}

@article{Itzykson:1983gb,
    author = "Itzykson, C. and Pearson, R. B. and Zuber, J. B.",
    title = "{Distribution of Zeros in Ising and Gauge Models}",
    reportNumber = "SACLAY-SPHT-83-40",
    doi = "10.1016/0550-3213(83)90499-6",
    journal = "Nucl. Phys. B",
    volume = "220",
    pages = "415--433",
    year = "1983"
}

@article{pisani1993lee,
  title={Lee-yang zeros and stokes phenomenon in a model with a wetting transition},
  author={Pisani, Claudio and Smith, Edgar R},
  journal={Journal of statistical physics},
  volume={72},
  number={1},
  pages={51--78},
  year={1993},
  publisher={Springer}
}

\end{document}